# Low-Complexity Distributed Radio Access Network Slicing: Algorithms and Experimental Results

Salvatore D'Oro, *Member, IEEE,* Francesco Restuccia, *Member, IEEE,*
Tommaso Melodia, *Fellow, IEEE,* and Sergio Palazzo, *Senior Member, IEEE*

***Abstract*—Radio access network (RAN) slicing is an effective methodology to dynamically allocate networking resources in 5G networks. One of the main challenges of RAN slicing is that it is provably an NP-Hard problem. For this reason, we design near-optimal low-complexity distributed RAN slicing algorithms. First, we model the slicing problem as a congestion game, and demonstrate that such game admits a unique *Nash equilibrium* (NE). Then, we evaluate the *Price of Anarchy* (PoA) of the NE, i.e., the efficiency of the NE as compared to the social optimum, and demonstrate that the PoA is upper-bounded by $3/2$. Next, we propose two fully-distributed algorithms that provably converge to the unique NE without revealing privacy-sensitive parameters from the slice tenants. Moreover, we introduce an adaptive pricing mechanism of the wireless resources to improve the network owner's profit. We evaluate the performance of our algorithms through simulations and an experimental testbed deployed on the Amazon EC2 cloud, both based on a real-world dataset of base stations from the OpenCellID project. Results conclude that our algorithms converge to the NE rapidly and achieve near-optimal performance, while our pricing mechanism effectively improves the profit of the network owner.**

## I. Introduction

Thanks to the ubiquitousness of modern smartphones and the rise of the Internet of Things (IoT), the number of mobile devices has seen an unprecedented expansion over the last few years. The latest report by Ericsson Mobility forecasts that the number of 5G subscriptions will exceed half a billion by the end of 2022, including more than 1.5 billion IoT devices equipped with cellular connections [1]. This massive increase in the number of connected devices will necessarily result in a staggering growth in cellular data traffic. For this reason, 5G cellular networks are expected to meet stringent requirements on ubiquitous connectivity, extremely low latency, and very high-rate data transfer [2–4].

What 5G systems are going to be has yet to be determined. However, there is wide consensus that 5G systems will single-handedly cater to a plethora of different services, such as automotive, mobile broadband, and tactile Internet, just to name a few. Each service will have its own networking requirements, which will ultimately require the network to be *polymorphic* and seamlessly adapt to different constraints. To address this challenging problem, the notion of *network slicing* [5–9] has been recently proposed, where the physical and computational resources of the network are seen as a unique "object" that can be "sliced" and "served" according to an entity's current needs. In this way, heterogeneous requirements can be served by the same infrastructure in a cost-effective manner, as different network slice instances can be orchestrated and configured according to the specific requirements of the slice tenants.

S. D'Oro, F. Restuccia and T. Melodia are with the Department of Electrical and Computer Engineering, Northeastern University, Boston, MA, 02115 USA e-mail: {s.doro, f.restuccia, t.melodia}@northeastern.edu.

S. Palazzo is with the Dipartimento of Ingegneria Elettrica, Elettronica e Informatica (DIEEI), University of Catania, Catania 92125 Italy. Email: sergio.palazzo@dieei.unict.it.

**Challenges.** Network slicing does not just provide better network performance, but it also enables realistic and profitable business models in the mobile network ecosystem. In this paper, we consider the problem of network slicing in the context of radio access networks (RANs), where telco operators (TOs) provide physical network resources to mobile virtual network operators (MVNOs), who periodically rent slices to provide cellular network access to mobile users (MUs). While a similar model is currently being successfully applied by Amazon Web Services or Microsoft Azure in the context of cloud services, slicing RAN resources is an intrinsically different problem, since (i) spectrum is a scarce resource for which over-provisioning is not possible; (ii) the network capacity is dynamic and heavily depends on the location of the MUs, among other factors; and (iii) the agreements with MVNOs usually impose stringent requirements on the Quality of Experience (QoE) perceived by the MUs.

The main issue in designing RAN slicing algorithms is that the most realistic problem formulations are provably NP-hard [10]. For this reason, most of existing work has focused only on the architectural aspects of network slicing [11, 12], with a limited focus on algorithmic aspects. Only very recently have a number of centralized algorithms for RAN slicing been proposed [10, 13]. Although these algorithms achieve optimality, they do not scale with the number of MVNOs and RAN resources. This issue calls for the design and analysis of *near-optimal, lower-complexity* algorithms for RAN slicing. Another crucial aspect is that optimality is often achieved to the detriment of privacy. Specifically, it is assumed that the TO is provided with complete information regarding the MVNOs' system parameters and preferences, for example, monetary budget, number and location of served MUs, business strategies, and so on. Instead, in realistic scenarios MVNOs may be reluctant to disclose such sensitive information to the TOs. Finally, centralized solutions do not consider the competitive and selfish behavior of modern MVNOs, who will dynamically adopt different strategies to maximize their own utility without considering the broader needs of the network as a whole. This implies that *distributed* algorithms are in general more desirable than centralized solutions, as they are more likely to be adopted by MVNOs in real-world RAN slicing scenarios.

**Contributions.** In this paper, we investigate the challenging and novel problem of designing privacy-preserving, low-complexity, near-optimal distributed algorithms for RAN network slicing, where the MVNOs selfishly compete with each other to acquire slices from the TOs while minimizing their cost. First, we first mathematically formulate this problem; then, we rely on the game-theoretical framework of congestion games [14–16] to effectively model and analyze the competition among MVNOs. Next, we demonstrate the existence and uniqueness of the *Nash equilibrium* (NE) associated to the game, and prove that its *Price of Anarchy* (PoA), i.e., the efficiency of the NE as compared to the social optimum, is limited to $^3/_2$. We formulate two distributed algorithms that preserve the privacy of the MVNOs and provably converge to the unique NE. Moreover, we propose a pricing algorithm based on a stochastic iterative mechanism that allows the TO to optimize its profit by adapting pricing policies to the current load on the resources being sliced. Finally, we extensively evaluate the performance of our algorithms through simulations and a practical testbed implemented on the Amazon EC2 cloud [17], both based on a real-world dataset of base stations (BSs) from the OpenCellID project [18]. Results demonstrate the effectiveness of the proposed approach, as our algorithms converge to the NE rapidly while achieving near-optimal performance.

**Paper Organization.** The paper is organized as follows. Section II introduces the network scenario, while Section III defines the RAN slicing problem. Sections IV to VII introduce and analyze the network slicing problem as a congestion game, and present the distributed and privacy-preserving algorithms that compute the NE for the MVNOs in such game. Pricing and profit issues are discussed in Section VII. Sections VIII and VIII-D present the numerical and experimental results, while Sections IX and X conclude the paper by discussing related work and drawing conclusions, respectively.

## II. Network Scenario and Problem Overview

In this paper, we consider the network scenario depicted in Figure 1, which is composed by a core network (CN) and a radio access network (RAN). The CN is connected to the Internet and is in charge of routing uplink and downlink traffic. Furthermore, the RAN is composed of multiple heterogeneous remote radio heads (RRHs). We assume that both the CN and the RAN are owned and managed by a single TO. However, our model can also be applied to the more general case of multiple infrastructure owners. Given that RRHs are geographically located at different areas of the network, the RAN can be divided into multiple *RAN clusters*, which group together RRHs that are close to each other.

In the considered scenario, the TO dynamically leases the network infrastructure to multiple MVNOs, who do not own any infrastructure and build virtual RANs to provide MUs with networking capabilities [6]. Since RAN clusters are geographically isolated and autonomous, slicing policies are enforced on each RAN cluster individually.

We consider the problem of generating autonomous *virtual* RANs on top of the resources provided by the physical RAN owned by the TO. A virtual RAN is thus referred to as a *network*

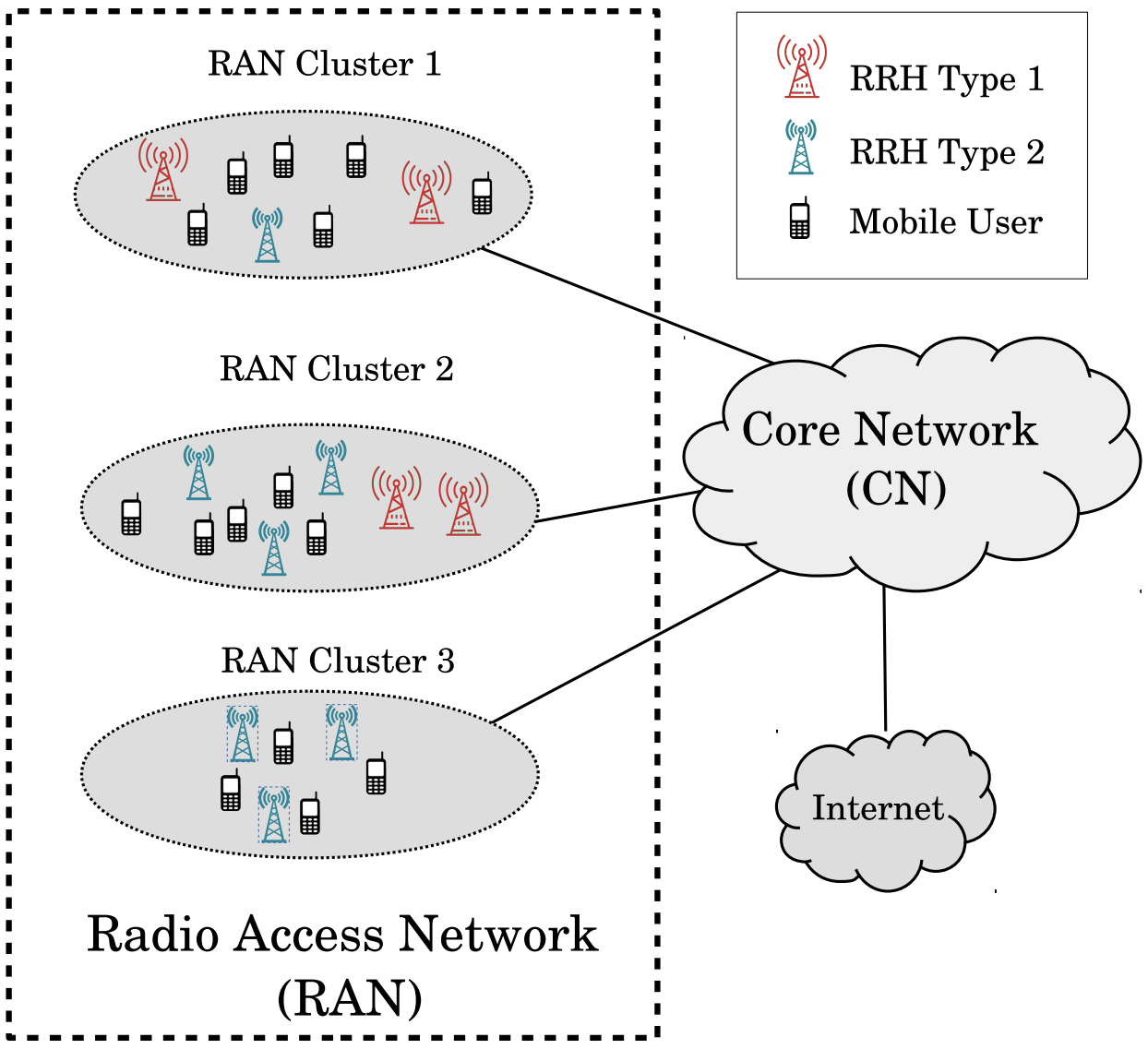

Figure 1. The considered 5G network scenario.

*slice*, since it represents a virtual "portion" of the physical RAN. Resources that may be assigned in each slice include available spectrum, antennas, computing resources, and so on. In this scenario, it is realistic to assume the MVNOs will need to provide the TO with some sort of payment based on the number and type of resources allocated to their slices. Thus, the slicing problem reduces to finding an allocation policy for each MVNO, to minimize the cost associated to its slice while ensuring that a predetermined number of MUs can be served. Within each slice, MVNOs are free to assign resources to the MUs according to their internal policies.

This problem can be approached in two different ways. In the first approach, which we call *centralized*, the TO collects the number of MUs that each MVNO would like to serve, referred to as *MVNO user load*. Then, the TO computes the slices by solving an optimization problem that takes the MVNO user loads into account. Although simple in nature, this approach suffers from a number of disadvantages. First, it requires the MVNOs to disclose their user load to the TO, which clearly raises privacy concerns as the MVNOs are usually interested in hiding such information from external entities. Second, since centralized approaches find the global optimum by mathematical optimization [10], they require significant computational resources as the network slicing problem becomes intractable as the number of MVNOs and resources increases [13]. Thus, since the MVNO user load is dynamic in nature, the centralized approach may fail to provide slices for the MVNOs before their user load changes again and a different slice is thus needed.

In this paper, we take a different approach and tackle the problem in a *distributed* way. Specifically, we design distributed slicing algorithms such that each MVNO computes its slice *without revealing information to the TO or other MVNOs*. Although such distributed solution is necessarily sub-optimal, we show that it is low-complexity (i.e., scalable) and approximates the optimum solution by a factor of $^3/_2$, a very small number. We use the framework of congestion games [14–16] to ensure that the MVNOs, although selfish and competing with each other, will nevertheless reach a solution (i.e., Nash equilibrium) that will satisfy each MVNO. In order to increase the TO's profit, we propose an adaptive resource pricing algorithm. Fig. 2 provides a roadmap of the following analytical sections.

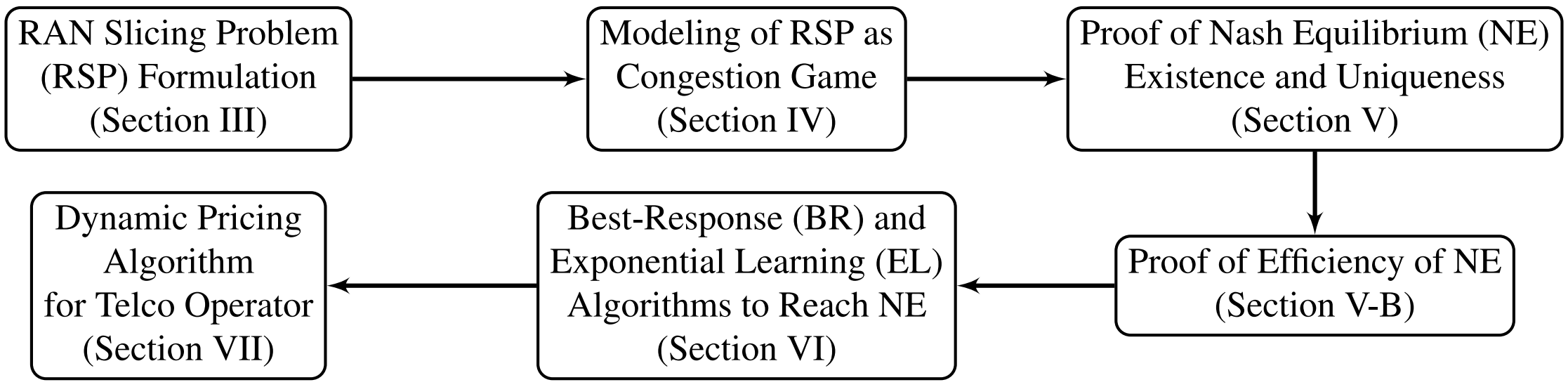

Figure 2. Roadmap to our theoretical analysis.

## III. RAN Slicing Problem Formulation

Let us define $\mathcal{M}$ and $C$ as the sets of the MVNOs and RAN clusters, respectively. We assume that a set $\mathcal{R}_c$ of $R$ heterogeneous RRHs are deployed for each cluster $c \in C$, and that RRHs can be classified into $L$ classes according to their performance and resource availability (e.g., macro, micro, pico and femto cells). Accordingly, the set of RRHs in RAN cluster $c$ is $\mathcal{R}_c = (\mathcal{R}_c^{(1)}, \mathcal{R}_c^{(2)}, \ldots, \mathcal{R}_c^{(L)})$. For each RRH $r \in \mathcal{R}_c$, we also assume that the TO fixes a leasing price, here denoted as $p_r$.

It is reasonable to assume that statistical information on the expected number of MUs and their position in each cluster may be obtained by exploiting real-time monitoring of the RAN [19]. Given such statistics are time-varying and depend on MUs mobility patterns, we assume that MVNOs change their slicing policy on regular basis. For these reasons, we assume time is discretized into slots, where $t = 1, 2, \ldots$ represents the slot index, and the number of MUs and their position in each cluster is assumed to vary at each slot.

Let $\Lambda_r$ be the maximum amount of resources of RRH $r$ available at each slot[1], and let $N_r$ denote the number of MUs that can be served by RRH $r$ without any degradation in QoE. In other words, if the number of MUs served by RRH $r$ is lower than $N_r$, then all MUs connected to $r$ enjoy high-quality communication. On the other hand, MUs would experience congestion and/or poor performance if their number is equal or greater than $N_r$.

The $N_r$ parameter is specific to a given RRH and depends on $\Lambda_r$, the desired minimum QoE level, the position of MUs and their number at a given slot. As an example, it is easy to observe that communication between distant RRHs and MUs is expected to result in more collisions and errors, which necessarily increases the number of re-transmissions and the transmission power. In general, the value of $N_r$ can be either fixed by hardware [22] or estimated by the TO by analyzing statistical information collected by using historical data such as transmission patterns and activities of MUs. For the sake of completeness, in the following we propose a methodology to estimate the value of $N_r$ that accounts for QoE requirements, resource availability, position and number of MUs. For the sake of simplicity, henceforth we will focus on a given cluster $c \in C$, and we will omit both $c$ and the slot index $t$. We summarize the most relevant system parameters in Table I.

### A. Derivation of $N_r$

Let us consider an MU communicating with RRH $r$. The received power at the MU's side can be denoted as follows:

$$p_r = k \left(\frac{d_0}{d_r}\right)^{\alpha} \tag{1}$$

where $k$ is a constant that accounts for transceiver design parameters such as antenna gains, channel characteristics, transmission power of the RRH and operating frequency; $d_0$ is a reference distance for far-field communication; $d_r$ is the distance between the MU and $r$; and $\alpha$ is the path loss coefficient [23]. Specifically, $\alpha = 1$ corresponds to the linear path loss model, $\alpha \leq 2$ models the free space or indoor path loss, and $\alpha > 3$ models urban and sub-urban scenarios.

Let us assume that the the RRH $r$ serves all MUs in its range with the same transmission power level. Accordingly, the Signal-to-Interference-plus-Noise Ratio (SINR) measured by each MU at distance $d_r$ from $r$ can be written as

$$\text{SINR}_r = \frac{p_r}{N + \tilde{N}_r p_r} \tag{2}$$

where $N$ is the channel noise power, $p_r$ is defined in (1), and $\tilde{N}_r$ represents the expected number of interfering MUs in the RRH coverage range [23]. Given that RRH $r$ can not allocate more than $\Lambda_r$ resources, the RRH may be modeled as an M/M/1 queue system with serving rate of $\Lambda_r$. To avoid queue starvation and enjoy the minimum SINR in (2), it follows that the arrival rate of MUs must not exceed $\Lambda_r$. In other words, $N_r$ should satisfy $N_r \leq \Lambda_r$.

Let $\mu \in [0, 1]$ such that $\mu \cdot N_r$ is the rate at which MUs request to access the available resources. By applying queuing theory [24], the expected number of the MUs in the system is derived $\frac{\mu \cdot N_r}{\Lambda_r - \mu N_r}$, while the expected number of interfering MUs in the RRH coverage range $\tilde{N}_r$ is derived as

$$\tilde{N}_r = \frac{\mu \cdot N_r}{\Lambda_r - \mu N_r} - 1 \tag{3}$$

[1] In our model, we do not assume any specific model for $\Lambda_r$, since it models a number of factors including computational, storage and communication resources. For example, $\Lambda_r$ might represent the number of OFDM symbols available in OFDM-based network architectures such as LTE [20], as well as spectrum or antennas [21].

By substituting (1) and (3) in (2), we obtain the maximum number $N_r$ of MUs that can be served simultaneously by $r$ while satisfying the minimum SINR constraint in (2) as follows:

$$N_r = \frac{\Lambda_r}{\mu} \cdot \frac{p_r(1 + \text{SINR}_r) - \text{SINR}_r N}{p_r \cdot (1 + 2 \cdot \text{SINR}_r) - \text{SINR}_r N} \tag{4}$$

where $p_r$ is a function of the distance between MUs and $r$ and is defined in (1).

*Remark* 1. From (4), we notice that (i) the maximum number of users that can be served by RRH $r$ is proportional to the available network resources $\Lambda_r$ and decreases as $\mu$ increases; and that (ii) $N_r$ decreases as the minimum SINR level and the distance $d_r$ increase. Furthermore, the variable $\mu$ can be varied at each slicing slot and be used to model the time-varying nature of cellular traffic at different part of the day. In Section VIII we show the impact of $\mu$ on different performance metrics such as the congestion level of the RRHs.

The above discussion provides an estimation of $N_r$ when MUs are located at a fixed distance $d_r$ from $r$. However, in real networks MUs are expected to change their position within the cluster continuously. Thus, statistical information on the MUs' mobility distribution can be used to provide a more accurate estimation of $N_r$. Specifically, we model the distance between MUs and RRH $r$ as a continuous random variable with probability density function $f_{D_r}(d)$ and $N_r$ in (4) can be rewritten as a function of $d_r$, *i.e.*, $N_r \rightarrow N_r(d_r)$. Accordingly, the expected maximum number of MUs that can be served by RRH $r$ can be computed as

$$N_r = \int_0^{+\infty} f_{D_r}(\tau) N_r(\tau) d\tau \tag{5}$$

The value of $N_r$ depends on the distribution $f_{D_r}(d)$ of MUs in the cluster, and might be considerably different if different values of $\mu, \text{SINR}_r, N, \Lambda_r$ are considered. Also, since RRHs are located at different positions of the network, the function $f_{D_r}(d)$ is expected to vary across different RRHs.

### *B. Definition of Slicing Policy*

We define as *slicing policy* a function determining how each MVNO allocates its MUs to the available RRHs in a given RAN cluster. More formally, a slicing policy for an MVNO $m$ is an *allocation vector* $\boldsymbol{\xi}_m = (\xi_{m,r})_{r\in\mathcal{R}} \in [0, n_m]$, where $\xi_{m,r}$ represents the number of MUs in the cluster that the MVNO $m$ expects to serve through RRH $r$.

Ultimately, the TO generates the global *allocation policy* $\boldsymbol{\xi} = (\boldsymbol{\xi}_m)_{m\in\mathcal{M}}$ by considering all the allocation vectors for each cluster. For any given allocation policy $\boldsymbol{\xi}$, the TO generates a corresponding network slice over the available RRHs. Specifically, the amount of resources at RRH $r \in \mathcal{R}$ provided to MVNO $m$ under allocation policy $\boldsymbol{\xi}$ is:

$$\sigma_{m,r} = s(\Lambda_r, \boldsymbol{\xi}), \tag{6}$$

where $\Lambda_r$ is the maximum amount of resources of RRH $r$, and $s(\cdot)$ is the *slicing function*. For example, a simple (yet effective), slicing function can be provided by the *proportional allocation function* [25]

$$\sigma_{m,r} = \frac{\xi_{m,r}}{\sum_{l\in\mathcal{M}} \xi_{l,r}} \Lambda_r. \tag{7}$$

Table I
SUMMARY OF NOTATION

| Variable | Description |
|---|---|
| $\mathcal{M},\ \mathcal{R},\ C$ | Sets of MVNOs, RRHs and RAN Clusters |
| $M, R$ | Number of MVNOs and RRHs |
| $\mathcal{R}_c$ | Set of the $L$ classes of RRHs in cluster $c$ |
| $N_r$ | Num. of MUs served by RRH $r \in \mathcal{R}$ with guaranteed QoE |
| $p_r$ | Leasing price for RRH $r \in \mathcal{R}$ |
| $t = 1, 2, \ldots$ | Slicing slot index |
| $n_{m,c}(t)$ | Expected number of MUs in cluster $c$ for MVNOs $m$ at slot $t$ |
| $\xi_{m,r}(t)$ | Number of MUs of MVNO $m$ served through RRH $r$ |
| $\Gamma_m(\boldsymbol{\xi}_m)$ | Set of RRHs in $\mathcal{R}$ selected by MVNO $m$ |
| $\Theta_r(\boldsymbol{\xi})$ | Set of MVNOs that have selected RRH $r$ under policy $\boldsymbol{\xi}$ |
| $\boldsymbol{\xi}_m(t)$ | Allocation vector for MVNO $m$ |
| $\boldsymbol{\xi}(t)$ | Allocation policy at slot $t$ |
| $\Lambda_r$ | Amount of resources available at RRH $r$ |
| $\sigma_{m,r}(\boldsymbol{\xi})$ | Available resources allocated to MVNO $m$ under allocation policy $\boldsymbol{\xi}$ |
| $\phi_r(\boldsymbol{\xi})$ | Congestion level on RRH $r \in \mathcal{R}$ under allocation policy $\boldsymbol{\xi}$ |
| $\pi_m^{\mathcal{P}}$ | Weighing parameter for MVNO $m$ |
| $c_m^{\mathcal{C}}(\boldsymbol{\xi}_m,\ \boldsymbol{\xi}_{-m})$ | Congestion cost for MVNO $m$ under allocation policy $\boldsymbol{\xi}$ |
| $c_m^{\mathcal{P}}(\boldsymbol{\xi}_m)$ | Monetary costs for MVNO $m$ under allocation policy $\boldsymbol{\xi}$ |
| $c_m(\boldsymbol{\xi}_m, \boldsymbol{\xi}_{-m})$ | Cost function for MVNO $m$ under allocation policy $\boldsymbol{\xi}$ |
| $\mathcal{G}$ | RAN slicing congestion game |
| $\Phi(\boldsymbol{\xi})$ | Potential function for the game $\mathcal{G}$ under allocation policy $\boldsymbol{\xi}$ |
| $\boldsymbol{\xi}^{\text{OPT}}, \boldsymbol{\xi}^{\text{NE}}$ | Optimal and Nash Equilibrium allocation policies |
| $C^{\text{NE}},\ C^{\text{OPT}}$ | Social welfares computed under the NE and optimal allocation policies |
| $\gamma_n$ | Step-size parameter used in the learning procedure in (25) |
| $\Pi(t)$ | Profit ot the TO at slot $t$ |

Since $\Lambda_r$ is a positive and finite real number, the relationship $\sum_{m\in\mathcal{M}} \sigma_{m,r} \leq \Lambda_r$ always holds for each $r \in \mathcal{R}$. For any $\boldsymbol{\xi}$, a slicing rule $\boldsymbol{\sigma}_r = (\sigma_{m,r})_{m\in\mathcal{M}}$ for each RRH $r \in \mathcal{R}$ is derived, and a *slicing policy* $\boldsymbol{\sigma} = (\boldsymbol{\sigma}_r)_{r\in\mathcal{R}}$ is obtained. For the sake of simplicity, we also define $\Gamma_m(\boldsymbol{\xi}_m)$ as the set of RRHs in $\mathcal{R}$ selected by MVNO $m$ in its slice, and $\Theta_r(\boldsymbol{\xi})$ as the set of MVNOs which have selected RRH $r$ under slicing policy $\boldsymbol{\xi}$. More formally, $\Gamma_m(\boldsymbol{\xi}_m) = \{r \in \mathcal{R} : \xi_{m,r} > 0, \xi_{m,r} \in \boldsymbol{\xi}_m\}$ and $\Theta_r(\boldsymbol{\xi}) = \{m \in \mathcal{M} : \xi_{m,r} > 0, \xi_{m,r} \in \boldsymbol{\xi}\}$.

## IV. RAN SLICING AS A CONGESTION GAME

When designing optimum slicing policies, MVNOs must also take into account aspects such as (i) the cost incurred when leasing RRHs; and (ii) QoS-related metrics such as (a) the distance of those RRHs from the MUs, (b) the congestion level on each RRH, and (c) the availability of resources at each RRH. Since MVNOs are assumed as selfish and interested in reaching their own optimum slicing policy, and given the RAN infrastructure is shared among MVNOs, slicing policies belonging to the same cluster will necessarily affect each other.

For these reasons, congestion games (CGs) [14–16] become the most natural choice to design and analyze efficient and distributed RAN slicing algorithms. Indeed, by using the theory of CGs, we are able to (i) demonstrate the existence and uniqueness of the *Nash equilibrium* (NE) associated to the CG; (ii) prove that its *Price of Anarchy* (PoA) is limited to $3/2$; and (iii) formulate two privacy-preserving distributed algorithms that provably converge to the unique NE.

We now formulate the RAN slicing problem as a congestion game and we define the slicing cost function used by MVNOs to select their optimum policy.

### A. Congestion Game Problem Formulation

Let us define and consider the following *weighted* congestion game (CG)

$$\mathcal{G} = (\mathcal{M}, (n_m)_{m\in\mathcal{M}}, \mathcal{R}, \mathcal{S}, (c_m)_{m\in\mathcal{M}}), \tag{8}$$

where $\mathcal{M}$ is the *player set* (i.e., the set of MVNOs), $n_m$ are the *weights* of the congestion game and represent the expected number of MUs in the cluster for MVNO $m$, $\mathcal{R}$ is the *resource set*, $\mathcal{S} = \prod_{m\in\mathcal{M}} \mathcal{S}_{m,c}$ is the *strategy space*, and $(c_m)_{m\in\mathcal{M}}$ is the *cost function set*, where $c_m$ will be defined in Section IV-B. A *strategy* for MVNO $m$ consists in the selection of the allocation vector $\boldsymbol{\xi}_m$. Hence, a strategy profile is represented by $\boldsymbol{\xi} = (\boldsymbol{\xi}_1, \boldsymbol{\xi}_2, \ldots, \boldsymbol{\xi}_M)$.

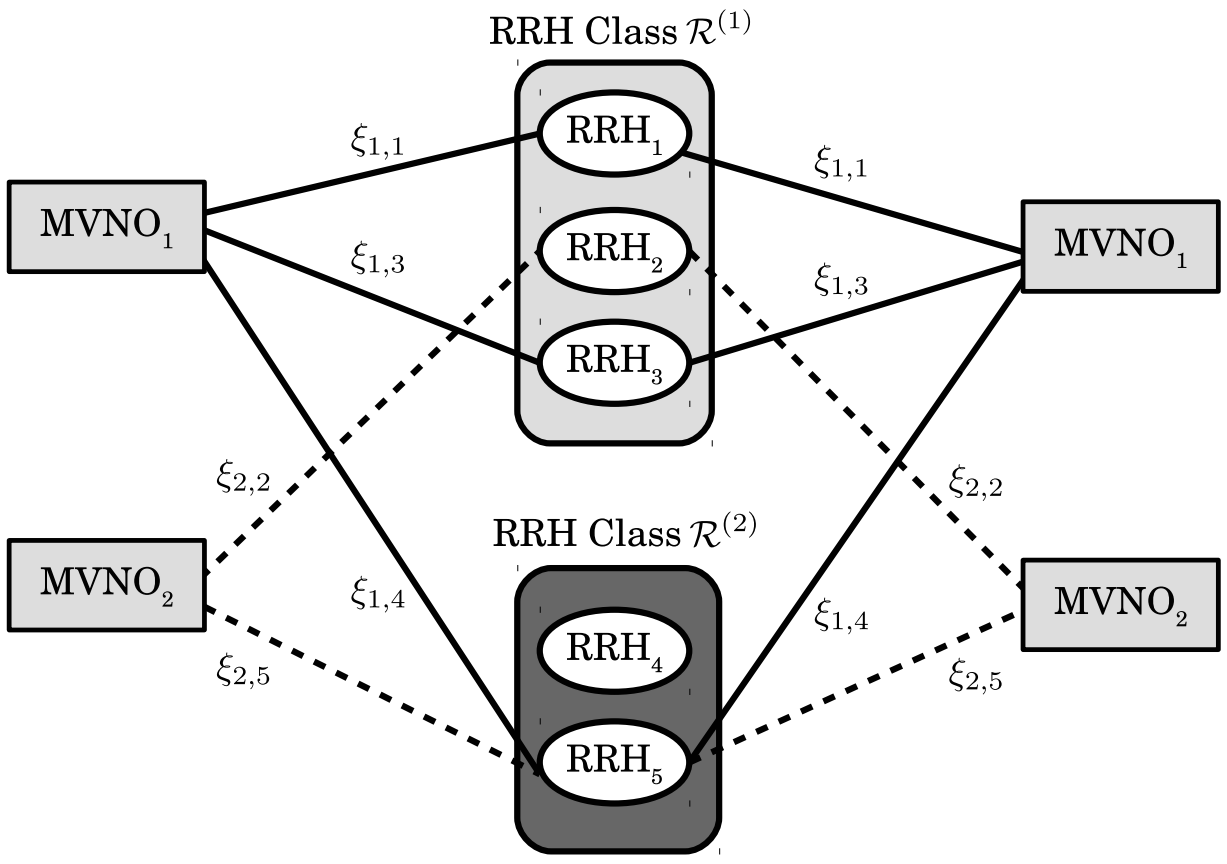

Figure 3. Example of the considered CG on a given cluster with $|\mathcal{M}| = 2$ MVNOs and $|\mathcal{R}| = 5$ heterogeneous RRHs.

Since the cost experienced by each MVNO depends on the RRHs they select, the congestion game $\mathcal{G}$ has to be weighted and to have *player-specific cost functions*. Furthermore, a slicing policy is derived by allocating different number of MUs to multiple RRHs in the cluster, implying that $\mathcal{G}$ has to be a CG with *splittable flows* [26]. To summarize, the slicing problem can be modeled as an atomic weighted CG with splittable flows on $R$ parallel links. Figure 3 shows a CG with two MVNOs. In this example, five RRHs are divided into two distinct RRH classes $\mathcal{R}^{(1)}$ and $\mathcal{R}^{(2)}$. Finally, $\xi_{m,r}$ represents the number of MUs allocated to RRHs $m$ by MVNO $m$.

### B. Definition of Slicing Cost Function

RRHs are heterogeneous in terms of type, position and available resources (i.e., spectrum and computational power). Thus, when selecting an RRH $r$, an MVNO $m$ will incur in a cost that depends on the following two aspects:

1) *RRH Congestion.* RRHs serving an excessive amount of MUs are expected to provide low QoS to their MUs. Instead, RRHs that serve fewer MUs might provide their served MUs with better quality communication. Therefore the RRH congestion should be taken into account. The *congestion level* on RRH $r$ under policy $\boldsymbol{\xi}$ can be measured as follows:

$$\phi_r(\boldsymbol{\xi}) = \frac{1}{N_r} \sum_{m\in\Theta_r(\boldsymbol{\xi})} \xi_{m,r}. \tag{9}$$

where $N_r$ has been defined in Section III-A and depends on several RRH-specific parameters such as the resource availability $\Lambda_r$ at RRH $r$ and the quality of the wireless channel. It is worth noting that the congestion cost not only depends on the allocation policy $\boldsymbol{\xi}_m$ of MVNO $m$, but it also depends on other MVNOs' allocation policies.

2) *Deployment Cost.* A second cost term is the *deployment cost* which MVNO $m$ pays to the TO to add RRH $r$ to its slice. Such cost depends on the price $p_r$ associated to RRH $r$. It models the revenue that the TO expects to receive when renting resources to MVNOs.

The above two aspects can be considered by defining the following resource- and MVNO-specific cost function

$$c_{m,r}(\boldsymbol{\xi}) = \phi_r(\boldsymbol{\xi}) + \pi_m^{\mathcal{P}} p_r \tag{10}$$

where the weight $\pi_m^{\mathcal{P}}$ is a resource- and MVNO-specific parameter that weighs the two terms in (10). $\pi_m^{\mathcal{P}}$ can be used to model a wide variety of network scenarios. For example, $\pi_m^{\mathcal{P}} = 0$ can be used to model a MVNO $m$ that cares about congestion only irrespective of the incurred deployment monetary cost.

The total cost incurred by MVNO $m$ under allocation policy $\boldsymbol{\xi}$ can be expressed as

$$c_m(\boldsymbol{\xi}_m, \boldsymbol{\xi}_{-m}) = \sum_{r\in\Theta_r(\boldsymbol{\xi})} \xi_{m,r}\, c_{m,r}(\boldsymbol{\xi}_m, \boldsymbol{\xi}_{-m}) \tag{11}$$

$$= c_m^{\mathcal{C}}(\boldsymbol{\xi}_m, \boldsymbol{\xi}_{-m}) + c_m^{\mathcal{P}}(\boldsymbol{\xi}_m) + c_m^{\mathcal{D}}(\boldsymbol{\xi}_m) \tag{12}$$

where

$$c_m^{\mathcal{C}}(\boldsymbol{\xi}_m, \boldsymbol{\xi}_{-m}) = \sum_{r\in\Phi_m(\boldsymbol{\xi}_m)} \xi_{m,c,r}\, \phi_r(\boldsymbol{\xi}_m, \boldsymbol{\xi}_{-m}) \tag{13}$$

$$c_m^{\mathcal{P}}(\boldsymbol{\xi}_m) = \pi_m^{\mathcal{P}} \sum_{r\in\Phi_m(\boldsymbol{\xi}_m)} \xi_{m,r}\, p_r \tag{14}$$

In the expressions above, (13) measures the overall congestion level experienced by all MUs served by the MVNO $m$ under the slicing policy $\boldsymbol{\xi} = (\boldsymbol{\xi}_m, \boldsymbol{\xi}_{-m})$; and (14) represents the total monetary deployment cost of the requested slice.

## V. Nash Equilibrium Analysis

In the previous section, we showed how the slicing problem can be modeled as a congestion game $\mathcal{G}$. In this section, we solve $\mathcal{G}$ by computing the Nash Equilibrium (NE) of the game, defined as follows:

**Definition 1.** A strategy profile $\boldsymbol{\xi} = (\boldsymbol{\xi}_m)_{m\in\mathcal{M}} \in \mathcal{S}$ is a NE for the congestion game $\mathcal{G}$ if, for all $m \in \mathcal{M}$, $(\boldsymbol{\xi}'_m, \boldsymbol{\xi}_{-m}) \in \mathcal{S}$ it holds that $c_m(\boldsymbol{\xi}_m, \boldsymbol{\xi}_{-m}) \leq c_m(\boldsymbol{\xi}'_m, \boldsymbol{\xi}_{-m})$, where $c_m(\cdot)$ is the cost function defined in (11).

Thus, a NE is a strategy profile such that no MVNO can further reduce its own cost function by unilaterally switching to another strategy.

In the following, we will focus our efforts to address the next four questions:

1) Does game $\mathcal{G}$ admit one or more NEs?
2) Is the NE efficient with respect to the optimal solution?
3) Can we design an efficient distributed algorithm that will help the MVNOs reach the NE without interacting with each other?

Question 1) will be answered in Section V-A, while questions 2) and 3) will be answered in Sections V-B and VI, respectively.

### A. Existence and Uniqueness of the NE

We first introduce the concept of *exact potential games* in Definition 2, and then we show that game $\mathcal{G}$ is an exact potential game in Proposition 3.

**Definition 2.** The game $\mathcal{G}$ is an exact potential game if there exists a function $\Phi : \mathcal{S} \rightarrow \mathbb{R}$ such that $c_m(\boldsymbol{\xi}_m, \boldsymbol{\xi}_{-m}) - c_m(\boldsymbol{\xi}'_m, \boldsymbol{\xi}_{-m}) = \Phi(\boldsymbol{\xi}_m, \boldsymbol{\xi}_{-m}) - \Phi(\boldsymbol{\xi}'_m, \boldsymbol{\xi}_{-m})$ for all $m \in \mathcal{M}$, $\boldsymbol{\xi}_m, \boldsymbol{\xi}'_m \in \mathcal{S}_m$, and $\boldsymbol{\xi}_{-m} \in \mathcal{S}_{-m}$, where $\mathcal{S}_{-m} = \prod_{k \in \mathcal{M}, k \neq m} \mathcal{S}_k$.

For each MVNO $m \in \mathcal{M}$ and RRH $r \in \mathcal{R}$, let $\eta_{m,r}$ be defined as follows:

$$\eta_{m,r} = \pi_m^{\mathcal{P}} p_r \tag{15}$$

**Proposition 3** (Potential Function and NE Uniqueness)**.** *The congestion game $\mathcal{G}$ is an exact potential game with potential function*

$$\Phi(\boldsymbol{\xi}) = \sum_{r \in \mathcal{R}} \sum_{m \in \mathcal{M}} \left( \frac{1}{N_r} \xi_{m,r}^2 + \eta_{m,r} \xi_{m,r} + \frac{1}{N_r} \sum_{k<m} \xi_{m,r} \xi_{k,r} \right). \tag{16}$$

*Furthermore, the congestion game $\mathcal{G}$ admits a unique NE.*

For a detailed proof see Appendix A.

### B. The Price of Anarchy (PoA)

In this section, we investigate the efficiency of the unique NE of the CG. Specifically, we will demonstrate that the *Price of Anarchy* (PoA) of the game is a small number. In particular, the PoA is a metric which measures the efficiency of the NE w.r.t. a social optimum solution, where the latter is computed as follows:

$$\boldsymbol{\xi}^{\text{OPT}} = \arg \min_{\boldsymbol{\xi} \in \mathcal{S}} \sum_{m \in \mathcal{M}} c_m(\boldsymbol{\xi}) \tag{17}$$

In our considered case, let $\boldsymbol{\xi}^{\text{NE}}$ be the unique NE of $\mathcal{G}$, the PoA can be expressed as

$$\text{PoA} = \frac{\sum_{m \in \mathcal{M}} c_m(\boldsymbol{\xi}^{\text{NE}})}{\sum_{m \in \mathcal{M}} c_m(\boldsymbol{\xi}^{\text{OPT}})} = \frac{C^{\text{NE}}}{C^{\text{OPT}}} \tag{18}$$

Since each MVNO aims at minimizing its cost function, it holds that PoA $\geq 1$. Specifically, PoA $= 1$ ensures the global optimality of the NE, instead, PoA $>> 1$ indicate poor efficiency of the NE. Even though many congestion games have been shown to have unbounded PoA, i.e., PoA $= +\infty$, in this work we show that the PoA is upper-bounded by $3/2$. Specifically, in Proposition 4 we first derive a general result on the PoA, then in Theorem 5 we derive a closed-form upper-bound which only depends on the number $M$ of MVNOs.

In line with [27], let $\beta$ be defined as follows

$$\beta = \sup_{r \in \mathcal{R}} \sup_{\boldsymbol{\xi}, \boldsymbol{\nu} \in \mathcal{S}} \left\{ \frac{\sum_{m \in \mathcal{M}} \xi_{m,r} \big(c_{m,r}(\boldsymbol{\xi}) - v_{m,r}(\boldsymbol{\xi})\big) + \nu_{m,r} \big(v_{m,r}(\boldsymbol{\xi}) - c_{m,r}(\boldsymbol{\nu})\big)}{\sum_{m \in \mathcal{M}} \xi_{m,r} c_{m,r}(\boldsymbol{\xi})} \right\} \tag{19}$$

where $v_{m,r}$ is the first-order partial derivative of $\xi_{m,r} c_{m,r}$ in (10), and is defined as

$$v_{m,r}(\boldsymbol{\xi}) = c_{m,r}(\boldsymbol{\xi}) + \frac{\xi_{m,r}}{N_r}. \tag{20}$$

**Proposition 4.** *The relationship $C^{\text{NE}} \leq \frac{1}{1-\beta} C^{\text{OPT}}$ holds, i.e.,* PoA $\leq \frac{1}{1-\beta}$.

Please refer to Appendix B for the complete proof.

**Theorem 5** (The PoA is upper bounded)**.** *The PoA is upper bounded by* $\text{PoA}(M) = \frac{3M+1}{2M+2}$. *That is, the PoA increases with the number $M$ of MVNOs, and is upper-bounded by* $\text{PoA}(+\infty) = 3/2$.

For the detailed proof see Appendix C.

It is worth noting that the result in Theorem 5 matches the result in [28], where it has been shown that the PoA for atomic splittable congestion games is bounded by $\frac{3M+1}{2M+2}$. However, their result applies to cost functions which are only resource-specific, i.e., $c_{m,r} = c_r$ for all $m \in \mathcal{M}$. In Theorem 5, instead, we consider cost functions such as those in (10) which are both player and resource-specific, thus extending the results in [28].

## VI. Distributed Algorithms for computing Nash Equilibrium

In this section, we propose two algorithms that provably converge to the unique NE of game $\mathcal{G}$. Specifically, in Section VI-1 we develop a *best response* approach to develop a mechanism that computes a NE by solving a convex quadratic programming (QP) problem. Section VI-2 will be devoted to the design of a learning mechanism which provably converges towards a NE and further reduces the complexity of the NE computation.

*1) Best Response approach:* We introduce the concept of *best response* (BR) functions and then prove that algorithms based on BR converge towards the unique NE.

The BR is a function that minimizes the cost of each player $m$ given the strategies $\boldsymbol{\xi}_{-m}$ of the other players. Specifically, the BR for player $m$ is defined as

$$\boldsymbol{\xi}_m^{\text{BR}} = \arg \min_{\boldsymbol{\xi}_m \in \mathcal{S}_m} c_m(\boldsymbol{\xi}_m, \boldsymbol{\xi}_{-m}) \tag{21}$$

An iterative algorithm where each player updates its current strategy according to (21) is called a *best response dynamics* (BRD). The following Proposition 6 holds.

**Proposition 6.** *The sequential BRD converges towards the unique NE of $\mathcal{G}$. Furthermore, the BR of each MVNO $m \in \mathcal{M}$ is unique.*

*Proof:* Since $\mathcal{G}$ admits a potential function, from (16) we have that any BR that minimizes the cost function of a given player also reduces the value of the potential $\Phi$. Therefore, since the potential is bounded and always non-negative, the sequential BRD is an *improvement path* [29] and will surely converge towards the unique NE. Finally, let us note that the cost functions in (11) are strictly convex in each player's strategy $\boldsymbol{\xi}_m$. Thus, the BR of each MVNO is unique. ■

Note that Proposition 6 does not guarantee that the convergence towards the unique NE is attained in a finite number of iterations. However, potential games with continuous strategy space possess the *approximate finite improvement path* (A-FIP) property, which ensures the convergence to an approximate NE in a finite amount of iterations. Even though we cannot guarantee the convergence in finite time to the NE, in Section VIII we will show that such convergence is attained in a limited number of iterations in many practical scenarios.

Proposition 6 shows that it is possible to converge towards the NE through iterative BRs. The BR in (21) for any MVNO $m$ and any adversarial strategy profile $\boldsymbol{\xi}_{-m}$ is obtained by solving

$$\min_{\boldsymbol{\xi}_m \in \mathcal{S}_m} \mathbf{f}_m \boldsymbol{\xi}_m + \frac{1}{2}\boldsymbol{\xi}_m^T \mathbf{Q} \boldsymbol{\xi}_m \quad (22)$$
$$\text{subject to} \sum_{r \in \mathcal{R}} \xi_{m,r} = n_m, \ \boldsymbol{\xi}_m \geq 0,$$

where $\mathbf{f}_m = (f_{m,i})_{i \in \mathcal{M}}$ and $\mathbf{Q} = \text{diag}\left(\frac{2}{N_r}\right)_{r \in \mathcal{R}}$ are a $M$-dimensional row vector and a $M \times M$ diagonal matrix, respectively. with

$$f_{m,r} = \frac{1}{N_r}\left[\phi_r(\boldsymbol{\xi}) - \xi_{m,r}\right] + \eta_{m,r} \quad (23)$$

and $\phi_r(\boldsymbol{\xi})$ and $\eta_{m,r}$ being defined in (9) and (15), respectively.

An algorithmic implementation of the sequential BRD is described in Algorithm 1.

**Algorithm 1** Sequential BRD

```
Input R; {(x_r, y_r), p_r, φ_r(ξ)}_{r∈R}; Output The unique NE of G;
while Convergence is not achieved do
    for each m ∈ M do sequentially
        ξ_m = (ξ_{m,r})_{r∈R} ← the unique solution of Problem (22);
    end for
end while
```

Problem (22) is a strictly convex quadratic problem with linear constraints. Accordingly, a solution can be computed in polynomial time (typically in $O(R^3)$, where $R$ is the number of available RRHs). Furthermore, if the congestion level $\phi_r(\boldsymbol{\xi})$ is publicly available, (23) guarantees that Problem (22) can be locally solved by each MVNO without any communication with the other MVNOs. Accordingly, Algorithm 1 computes a NE in a distributed fashion.

*2) A low-complexity learning approach:* In the previous section, we have shown that the BR of each MVNO can be computed by solving a QP problem, which is generally solved in polynomial time $\approx O(R^3)$. However, if the number $R$ of RRHs is large, to compute a solution to Problem (22) would anyway require a non-negligible amount of time. Motivated by this latter discussion, in this section we provide another approach that builds on learning theory to provide a low-complexity approach to compute the unique NE of $\mathcal{G}$.

Let us introduce the following *exponential learning scheme*

$$\begin{cases} z_{m,r}[n+1] = z_{m,r}[n] - \gamma_n v_{m,r}(\boldsymbol{\xi}[n]) \\ \xi_{m,r}[n+1] = n_m \frac{e^{z_{m,r}[n+1]}}{\sum_{k \in \mathcal{R}} e^{z_{m,k}[n+1]}} \end{cases} \quad (24)$$

where $\gamma_n$ is the *step-size*, $v_{m,r}$ is defined in (20) and $n$ is the iteration indicator. Intuitively, the exponential map in (24) generates values of the allocation variable $\xi_{m,r}$ which always lie on the boundary of the $(R-1)$–simplex $\mathcal{S}_m$. In Proposition 7, we show that (24) converges to the unique NE of $\mathcal{G}$.

**Proposition 7.** *Let $\boldsymbol{\xi}^* \in \mathcal{S}$ be the unique NE of $\mathcal{G}$. If the step-size $\gamma_n$ are chosen such that $\sum_{n=1}^{+\infty} \gamma_n^2 < \sum_{n=1}^{+\infty} \gamma_n = +\infty$, the exponential learning scheme in* (24) *always converges towards $\boldsymbol{\xi}^*$ from any starting point $\boldsymbol{\xi}(0) \in \mathcal{S}$. A suitable choice of $\gamma_n$ is $\gamma_n = \frac{1}{n^\beta}$, with $\beta \in (0.5, 1]$.*

*Proof:* Let us first derive the continuous time version of (24) as follows:

$$\begin{cases} \dot{z}_{m,r} = v_{m,r}(\boldsymbol{\xi}) \\ \xi_{m,r} = n_m \frac{e^{z_{m,r}}}{\sum_{k \in \mathcal{R}} e^{z_{m,k}}} \end{cases} \quad (25)$$

From the strict convexity of $c_m$, it follows that $v_{m,r}$ is continuously differentiable, and thus Lipschitz-continuous. Since $v_{m,r}$ is always bounded, (25) admits a unique solution for any initial condition $\boldsymbol{\xi}(0)$. From (16), we have that the potential function $\Phi$ is strictly convex, which implies that $\Phi$ is also star-convex with respect to the unique NE $\boldsymbol{\xi}^*$. Accordingly, the theory of [30] ensures that the unique solution of (25) also corresponds to the unique NE $\boldsymbol{\xi}^*$, and (24) converges to $\boldsymbol{\xi}^*$ if $\sum_{n=1}^{+\infty} \gamma_n^2 < +\infty$ and $\sum_{n=1}^{+\infty} \gamma_n = +\infty$. Finally, we have that $\gamma_n = \frac{1}{t^\beta}$ satisfies the above conditions. ■

The algorithmic implementation of (24) is described in Algorithm 2.

**Algorithm 2** Exponential Learning Scheme

```
Input R; {(x_r, y_r), p_r, φ_r(ξ)}_{r∈R}; Output The unique NE of G;
Set (v_{m,r})_{r∈R,m∈M} = 0; (z_{m,r})_{r∈R,m∈M} = 0;
while Convergence is not achieved do
    for each m ∈ M do simultaneously
        ξ_{m,r} ← n_m e^{z_{m,r}} / Σ_{k∈R} e^{z_{m,k}};
        z_{m,r} ← z_{m,r} − γ_n v_{m,r}(ξ);
    end for
end while
```

As already discussed at the end of Section VI-1, Algorithm 2 can be implemented in a distributed fashion by only making the congestion level $\phi_r(\boldsymbol{\xi})$ on each RRH publicly available. Furthermore, in opposition to the BR-based mechanism in Section VI-1, lines 5, 6 in Algorithm 2 have complexity $O(1)$. Since both $\xi_{m,r}$ and $z_{m,r}$ have to be computed for each RRH in $\mathcal{R}$ and MVNO in $\mathcal{M}$, the overall per-iteration computational complexity is $O(RM)$, i.e., each iteration of Algorithm 2 can be computed in linear time.

In Proposition 7, we have shown that the learning scheme in (24), and thus Algorithm 2, converges to the unique NE of the game if some conditions on the step-size $\gamma_n$ are satisfied. As an example, a suitable choice of the step-size is $\gamma_n = \frac{1}{n^\beta}$. Though this latter setting guarantees the convergence of the learning scheme, it might generate slow-convergent dynamics. To overcome this issue, it has been shown in [31] that the convergence process can be improved by using fixed values of the step-size. Although the convergence for this latter approach cannot be theoretically proven, using a fixed step-size still allows Algorithm 2 to converge to the NE with high probability. Accordingly, in this paper we will only consider fixed step-size rules, and we refer the interested readers to [31] for a more detailed discussion on the impact of different step-size rules on the convergence of the exponential learning scheme (24).

## VII. Pricing Policies

In the previous sections, we have shown that the developed algorithm does not require perfect knowledge with respect to MVNOs' parameters, and effectively preserves the privacy of all MVNOs. In addition, although the proposed algorithm has limited complexity, we have proven that close-to-optimal solutions can be obtained with no need to run complex and time-consuming centralized approaches at the TO side. Despite the above important properties, another relevant issue that has not yet been considered is related to pricing policies enforced by the TO and their impact on the achievable profit.

It is worth noting that the set of MVNOs, the availability of RRHs and the position and number of MUs in each cluster vary at each slot $t$. In such a dynamic scenario, it is straightforward to show that static pricing policies could fail to provide high profit to the TO. On the contrary, adaptive pricing policies are more suited to deal with time-varying systems such as the one we are considering in this paper. For this reason, in this section we focus on adaptive pricing policies, and we propose a stochastic-based algorithm that exploits past observations to adapt the pricing policy at future slots.

Let us focus on a given cluster $c$. Furthermore, let $\boldsymbol{\xi}^*(t) = (\xi^*_{m,r}(t))_{m\in\mathcal{M},r\in\mathcal{R}}$ be the unique NE of $\mathcal{G}$ at slot $t$, and let $n_r(t)$ be the number of MUs served through RRH $r$ at the NE. We have that

$$n_r(t) = \sum_{m\in\mathcal{M}} \xi^*_{m,r}(t). \tag{26}$$

The profit of the TO is defined as

$$\Pi(t) = \sum_{r\in\mathcal{R}} p_r(t)n_r(t) - C(n_r(t)), \tag{27}$$

where $C(n_r(t))$ is the cost experienced by the TO to manage $n_r(t)$ MUs at RRH $r$.

The allocation variables $\xi^*_{m,r}$ are obtained by iteratively executing either Algorithm 1 or Algorithm 2. Unfortunately, there is no closed-form for those variables, which makes it hard to predict the behavior of the MVNOs, and the actual allocation of the MUs. Also, $\xi^*_{m,r}$ depends on the weighting parameters $\pi^{\mathcal{P}}_m$. Those values are not known by the TO and might vary at each slot $t$. Therefore, an adaptive pricing policy should be considered such that the price $p_r$ can be updated at each slot $t$.

In this paper, we propose the following stochastic-based approximation pricing mechanism

$$p_r(t+1) = p_r(t) + \sigma\left[n_r(t) - n_r(t-1)\right], \tag{28}$$

where $\sigma > 0$ is a fixed step-size used to weigh the two terms in the stochastic procedure. The proposed pricing scheme in (28) works as follows. If $n_r(t) > n_r(t-1)$, i.e., the number $n_r$ of MUs served through RRH $r$ has been increased in the last slot, then the price at the next slot is increased as well, i.e., $p_r(t+1) > p_r(t)$. Otherwise, if $n_r(t) < n_r(t-1)$, the price is decreased, and $p_r(t+1) < p_r(t)$.

Furthermore, since the achieved profit must always be non-negative, the condition $p_r(t+1) \geq C(n_r(t))$ has to be satisfied for each RRH $r$. Since the stochastic procedure in (28) might generate a price $p_r(t+1)$ that violates the above constraint, a minimum value $p_r(t+1) = C(n_r(t))$ is considered at each iteration of (28).

## VIII. Performance Evaluation

In this section, we investigate on the performance of the proposed congestion game-based slicing mechanism through numerical simulation and experimental results. To emulate a realistic network setup, in our simulations we have extracted a cluster $\mathcal{R}^{(\mathrm{FULL})}$ of multiple RRHs deployed in Boston, MA (USA) from the OpenCellID database [32]. A cluster consisting of 100 RRHs, and the position of each RRH is illustrated in Fig. 4. We assume that the density of wireless devices in the considered cluster equals the population density in Boston and is set to 5000 devices/km$^2$ [33]. Furthermore, we consider the case where a set $\mathcal{M}$ of $M = 20$ MVNOs are willing to competitively deploy RAN slices to provide MUs with wireless access to the internet. We assume that each MVNO $m$ serves the same number $n_m$ of devices. Hence, the density of wireless devices served by each MVNO is $5000/M$ devices/km$^2$.

We assume that all RRHs are LTE base stations operating at 2.4GHz. The path-loss exponent in (1) is set to $\alpha = 3$, e.g., urban scenario, and the channel noise power is $N = -174$dBm/Hz [31]. We consider isotropic antennas with antenna gain equal to 3dBi and a reference distance $d_0 = 1$m [34]. Thus, the $k$ parameter in (1) is $k = 9.89 \cdot 10^{-5}$ [23].

We assume that the amount of available resources at each RRH $r$, i.e., $\Lambda_r$, is represented by the number of LTE Resource Elements (RE) within an LTE Resource Block (RB). Specifically, we assume that each RB consists of $s_r = 7$ OFDM symbols transmitted over $c_r = 12$ subcarriers for all $r \in \mathcal{R}$. Note that the number of available RBs, and thus $\Lambda_r$ depends on the LTE channel bandwidth. Specifically, we have that $\Lambda_r = N_{\mathrm{RB}} \cdot s_r \cdot c_r$, where $N_{\mathrm{RB}} \in \{25, 50, 100\}$ represents the number of available RBs in the LTE system when the bandwidth is set to 5, 10, and 20 MHz, respectively. Unless otherwise stated, we assume $b = 20$MHz. That is, $\Lambda_r = 100$ REs are available for MU transmission at each RRH. Furthermore, to investigate the impact of different minimum Quality of Experience (QoE) levels on the performance of the slicing mechanism, in our simulations we consider three different minimum SINR requirements such that $\mathrm{SINR}_r \in \{-5, 0, 5\}$ dB.

The price $p_r$ associated to each RRH $r \in \mathcal{R}^{(\mathrm{FULL})}$ is assumed to be generated according to a normal distribution with mean value $\mu_p = 10$ Price Units (PU) and standard deviation $\sigma_p = 4$ PU. Instead, the weight $\pi^{\mathcal{P}}_m$ in (10) and the access rate $\mu$ of MUs in (4) are assumed to be uniformly distributed and are randomly generated at each simulation run. More in detail, $\pi^{\mathcal{P}}_m$ takes value in $[0, \pi^{\mathcal{P}}]$, with $\pi^{\mathcal{P}} = 5 \cdot 10^{-4}$, while $\mu \in [0, 1]$.

To investigate the impact of the number $R$ of RRHs deployed in the network on the achievable performance of the network, for any given $R \in \mathbb{Z}$, in our simulations we generate a subset $\mathcal{R} \subseteq \mathcal{R}^{(\mathrm{FULL})}$ of $R$ RRHs which are randomly picked from $\mathcal{R}^{(\mathrm{FULL})}$. The results presented in the following are averaged over 2000 independent simulation runs. The 95% confidence intervals are not shown when below 2% of the average.

### A. PoA and Convergence Speed Analysis

We first investigate the performance of the proposed RAN slicing algorithms as compared to an algorithm computing the social optimum as discussed in Section V-B. In Fig. 5, we show the PoA of the proposed slicing solution as a function of the ratio $R/M$ and for different values of $N_{\text{RB}}$ when $\mu = 0.8$. The figure concludes that the Price of Anarchy (PoA) is small and approximately 1 in all the considered cases, *i.e.*, the proposed solution provides near-optimal slicing of the network resources even though the cluster consists of approximately 100 RRHs. Furthermore, Fig. 5 shows that the proposed solution achieves better performance in terms of optimality gap when a small number of RBs is considered.

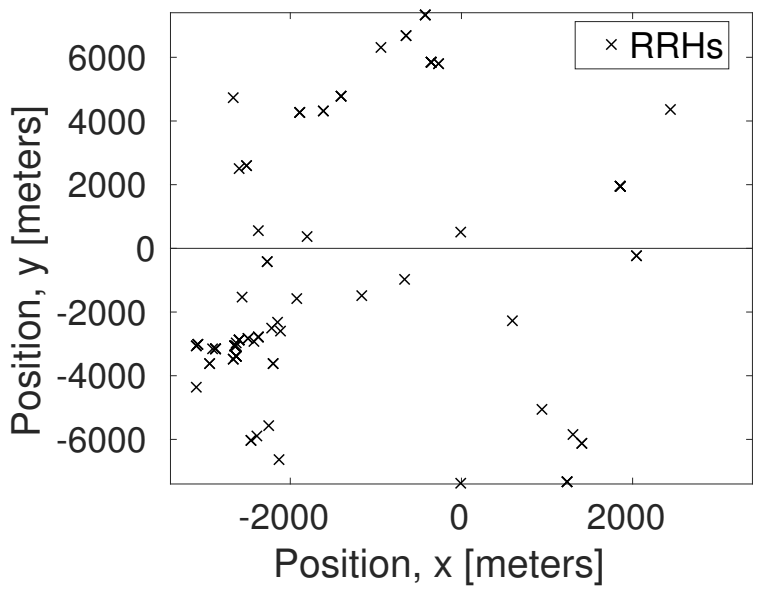

Figure 4. The considered cluster with the position of 100 RRHs taken from the OpenCellID database.

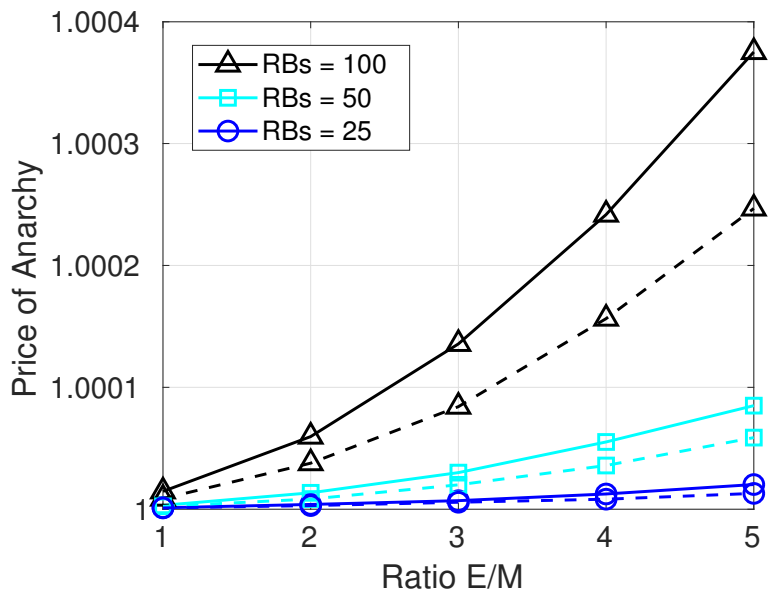

Figure 5. The PoA of the proposed slicing solution as a function of the ratio $R/M$ for different number $N_{\text{RB}}$ of RBs and minimum SINR requirement (Solid lines: SINR $= -5$dB; Dashed lines: SINR $= 5$dB).

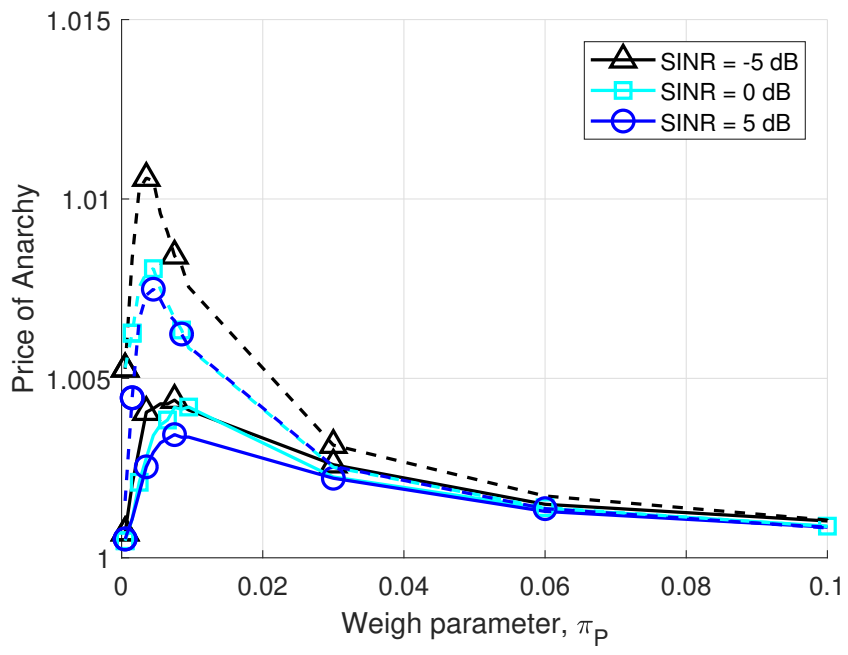

Figure 6. The PoA of the proposed slicing solution as a function of $\pi_m^{\mathcal{P}}$ for different values of the minimum SINR requirement and number of RRHs (Solid lines: $R = 20$; Dashed lines: $R = 50$).

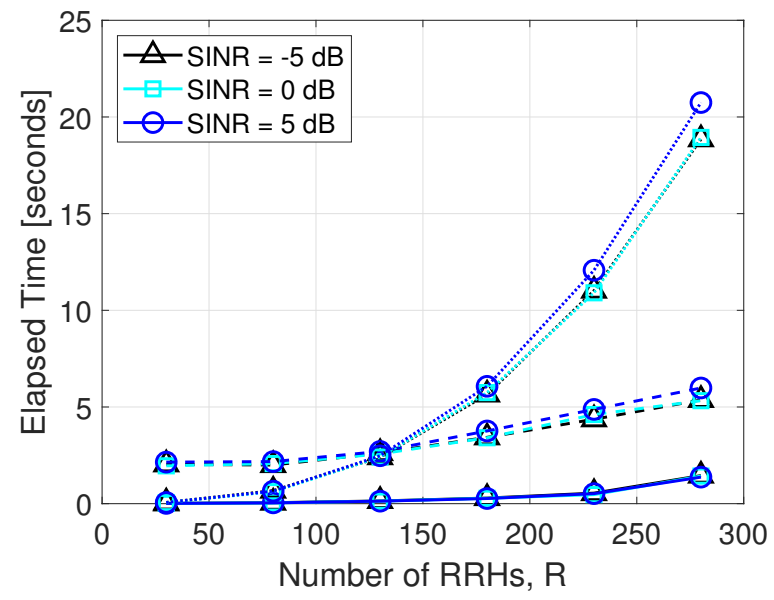

Figure 7. Execution time of several solutions for different minimum SINR requirements (Solid lines: Algorithm 2; Dashed lines: Algorithm 1; Dotted lines: Centralized optimal solution).

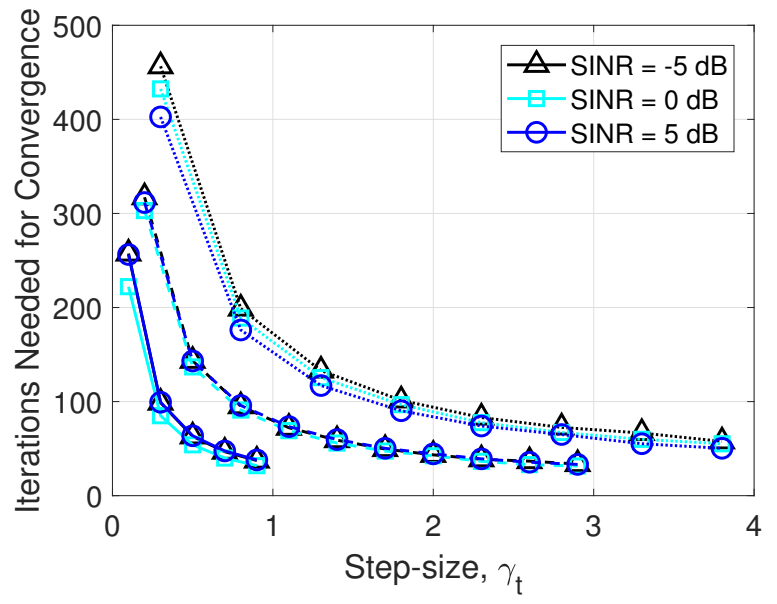

Figure 8. Number of iterations needed by Algorithm 2 to reach the NE as a function of $\gamma_n$ for different minimum SINR requirements and number of RRHs (Solid lines: $R = 20$; Dashed lines: $R = 50$; Dotted lines: $R = 100$).

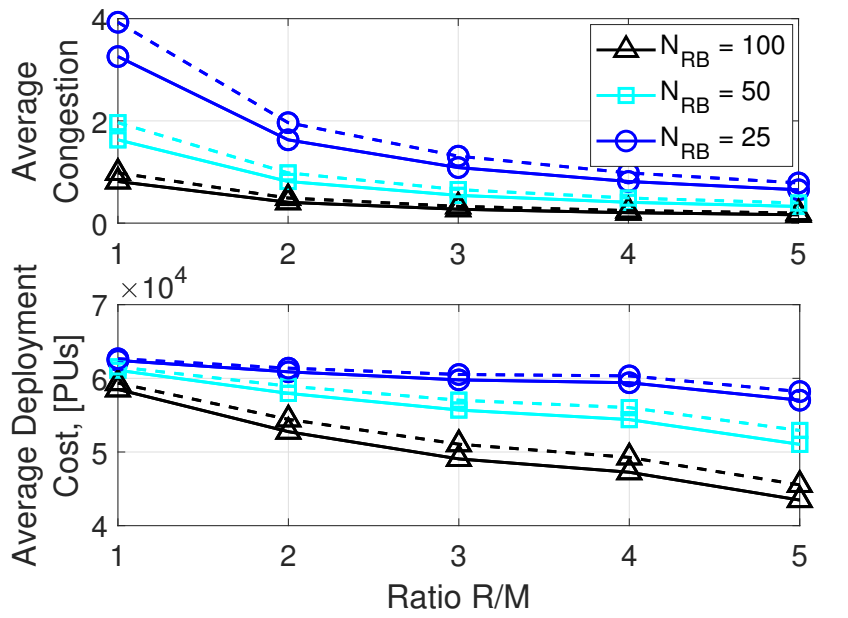

Figure 9. Average congestion level on each RRH and deployment cost of a slice as a function of the ratio $R/M$ for different number of RBs and minimum SINR requirements (Solid lines: SINR $= -5$dB; Dashed lines: SINR $= 5$dB).

Fig. 6 shows the impact of the weighing parameter $\pi_m^{\mathcal{P}}$ on the cost function (11) of the PoA. The weight $\pi_m^{\mathcal{P}}$ is uniformly distributed in $[0, \pi^{\mathcal{P}}]$. Accordingly, in Fig. 6 we let the upper-bound $\pi^{\mathcal{P}}$ vary for different values of the minimum SINR requirement. Fig. 6 clearly shows the existence of two distinct regions;the first region is associated to small values of $\pi^{\mathcal{P}}$, *i.e.*, those network scenarios where MVNOs primarily aim at minimizing the congestion cost in (11). In this case, the PoA increases until a maximum value is attained. Instead, when higher values of $\pi^{\mathcal{P}}$ are considered, *i.e.*, the deployment cost associated to each slice is no more negligible, the PoA asymptotically decreases to one until near-optimality is achieved.

Fig. 7 compares the execution time of the solution proposed in this paper. Specifically, we compare the time needed by a centralized algorithm to compute an optimal solution, and that needed by Algorithms 1 and 2 to compute a NE. The results conclude that the centralized algorithm (dotted lines) suffers from severe complexity issues when dense networks are considered. On the contrary, both the BRD (dashed lines) and learning (solid) algorithms show faster convergence rate. Furthermore, Algorithm 1 still requires to solve the QP problem (22), which usually has polynomial computational complexity. Instead, Algorithm 2 has linear per-iteration and per-user complexity, which results in faster convergence towards the NE.

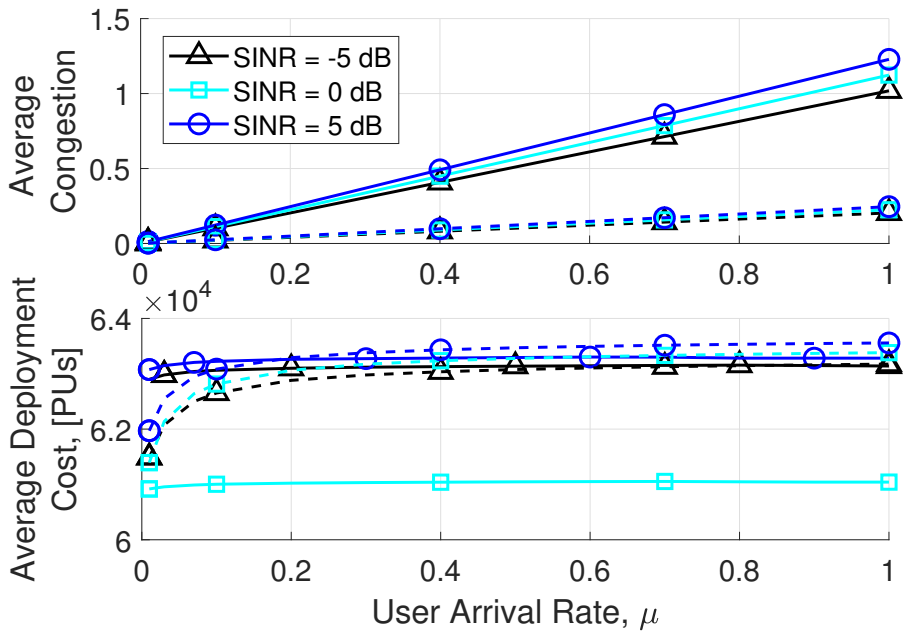

Figure 10. Average congestion level on each RRH and deployment cost as a function of $\mu$ for different values of the minimum SINR requirement and number of RRHs (Solid lines: $R = 20$; Dashed lines: $R = 100$).

To investigate the convergence speed of Algorithm 2 under the fixed step-size rule, in Fig. 8 is shown the number of iterations needed by the learning algorithm to converge to the NE as a function of the value of the step-size parameter for different minimum SINR requirements and values of the number $R$ of RRHs. It is worth noting that the higher the value of the step-size, the higher the convergence speed of the algorithm. Furthermore, the convergence of Algorithm 2 is slowed down when the number $R$ of RRHs is large. However, by increasing the value of the fixed step-size, the convergence speed is reduced even in those scenarios where the number of RRHs and MVNOs is high.

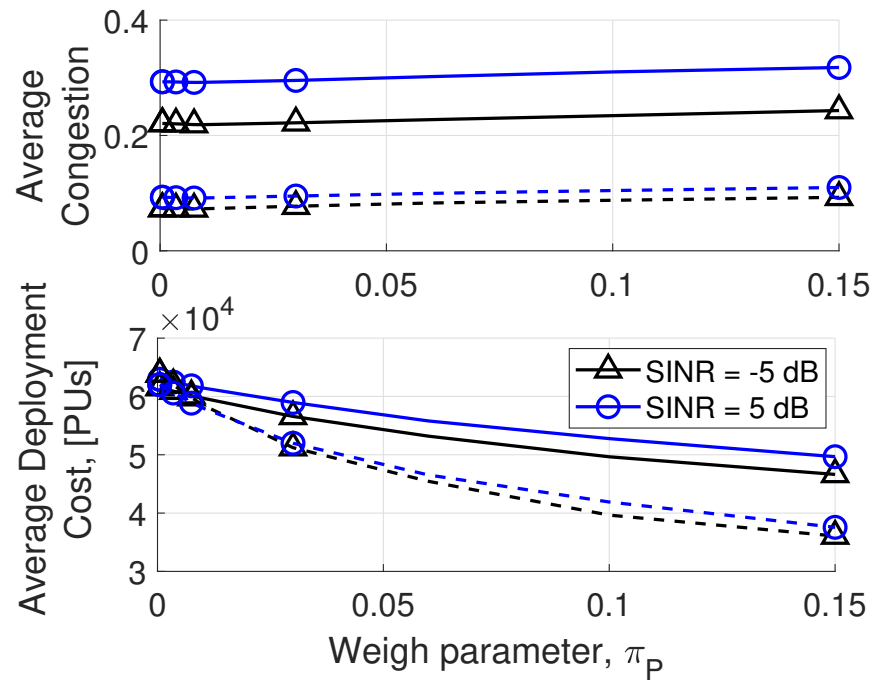

Figure 11. Average congestion level on each RRH and deployment cost as a function of $\pi_m^{\mathcal{P}}$ for different values of the minimum SINR requirement and number of RRHs (Solid lines: $R = 20$; Dashed lines: $R = 50$).

### *B. Congestion and Deployment Costs Analysis*

Fig. 9 shows the impact of the ratio $R/M$ on both the average congestion and the deployment cost of the network for different values of the number of RBs and the minimum SINR requirement when $\mu = 0.8$ and $\pi^{\mathcal{P}} = 5 \cdot 10^{-}2$. Specifically, the congestion at each RRH is evaluated as in (9). Instead, the deployment cost of each MVNO $m \in \mathcal{M}$ is computed according to (14) and then normalized by the corresponding weight $\pi_m^{\mathcal{P}}$ to obtain the actual monetary cost. The figure concludes that the congestion is decreasing as a function of $R/M$. Intuitively, this is because the deployment of additional RRHs on the network allows the MVNOs to generate RAN slices which contain different RRHs, thus inevitably reducing the congestion level on those RRHs. Furthermore, it is shown that the deployment cost decreases as well when a larger number of RRHs are deployed. In this case, MVNOs can select those RRHs with a low price $p_r$, thus reducing the overall monetary expenditure. Fig. 9 also shows that when SINR = 5dB (dashed lines), both the congestion level and the deployment cost of the network are higher as compared to the case where SINR = −5dB (solid lines). In other words, higher values of the minimum SINR requirements reduce the maximum number of MUs that can be served by each RRH, which, on average, leads to higher values of congestion and deployment cost.

From (4), the number $N_r$ of MUs that can be served with high QoE by each RRH decreases as the access rate $\mu$ of MUs increases. As a consequence, the congestion on each RRH increases due to the high traffic demand generated by MUs. This phenomenon is depicted in Fig. 10 where we show the average congestion level on each RRH and deployment cost of a slice as a function of the access rate $\mu$ for different values of the minimum SINR requirement and $R$. Intuitively, the congestion increases when both $\mu$ and the minimum SINR level are high, and a small number of RRHs is deployed. Accordingly, to support the high traffic demand, MVNOs add more RRHs to their slice, which increases the deployment cost of each slice.

The impact of $\pi_m^{\mathcal{P}}$ on the congestion level of the sliced network is investigated in Fig. 11, where we show the average congestion level and deployment cost as a function of $\pi^{\mathcal{P}}$ for different values of the minimum SINR requirement and number of RRHs when $\mu = 0.8$. The lowest values of congestion are achieved when $\pi^{\mathcal{P}} = 0$. In this case, from (11) we have that the proposed solution computes a NE that individually minimizes the congestion level on each slice. Accordingly, such a setting produces low congested slices. On the contrary, the average congestion increases when $\pi^{\mathcal{P}}$ increase as well. As expected, the average congestion when only few RRHs are available (i.e., dotted lines) is higher than the case where a high number $R$ of RRHs is deployed on the network (i.e., dashed lines). Fig. 11 also shows that the deployment cost of the slices decreases as $\pi^{\mathcal{P}}$ increases. Such a result stems from the fact that, by increasing the weight parameter, MVNOs are more affected by deployment costs and converge towards more conservative slicing strategies to avoid high monetary expenses.

### *C. Profit Analysis*

In this section, we investigate the profit achieved by the TO and we compare different pricing policies. Specifically, we consider three different pricing schemes as follows:

- *Uniform Pricing:* all the RRHs in $\mathcal{R}$ are equally priced, i.e., a fixed price equal to $p_r = \mu_p$ is enforced on each RRH $r \in \mathcal{R}$;
- *Weighted Pricing:* the price enforced on each RRH is proportional to $N_r$. Specifically, the price $p_r$ for the RRH $r \in \mathcal{R}$ is set to $p_r = \mu_p \frac{N_r}{\max_{e \in \mathcal{R}} N_e}$. Under this pricing policy, the TO fixes higher prices to those RRHs which can serve a higher number $N_r$ of devices simultaneously;
- *Adaptive Pricing:* this policy has been introduced and discussed in Section VII. Intuitively, at each slot $t$, the price of RRH $r$ is updated according to (28) by considering the congestion level on $r$ at slot $t-1$. From (28), it can be shown that the adaptive scheme will eventually converge towards a stable pricing policy when $t \longrightarrow \infty$.

The profit achieved by the TO is shown in Fig. 12 as a function of the mean value $\mu_p$ for different pricing schemes. Since the proposed adaptive pricing model updates the price of the RRHs at each slicing slot $t$, in Fig. 12 we show the profit achieved by the TO when $t \longrightarrow \infty$. As expected, the profit achieved by the TO increases when the value of $\mu_p$ increases as well, and it is shown that the proposed pricing model outperforms the other mechanisms.

### *D. Experimental Results*

To understand the impact of our system in real-world deployments, Fig. 13 shows the experimental results obtained by executing the learning algorithm on a testbed deployed on the Amazon Elastic Computing (EC2) cloud service. Specifically, we have deployed the TO's code on a *t2.micro* instance, having 1 virtual CPU, one 1 GByte of RAM with 3.3GHz clock speed. We have deployed the code implementing the MVNO's slicing policy learning algorithm on an iMac desktop computer located on Northeastern University campus. As we can see from

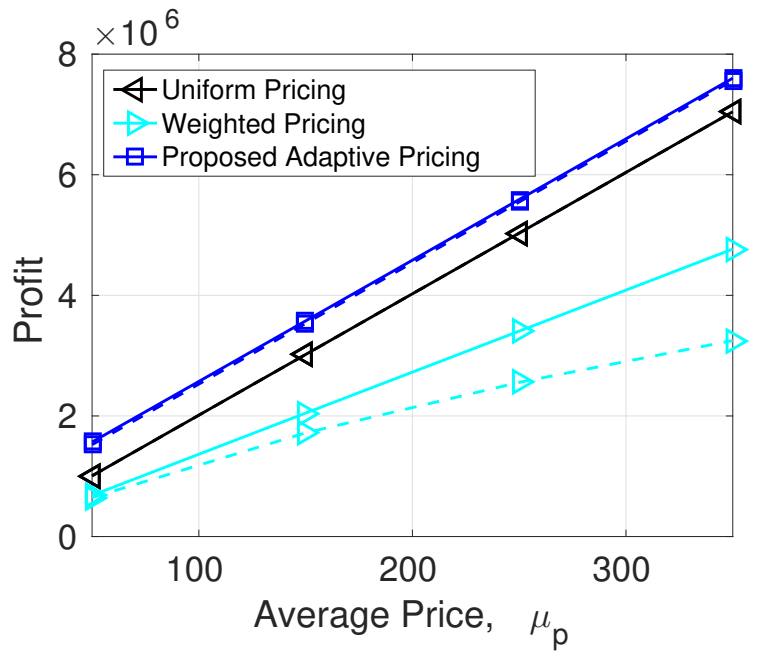

Figure 12. Profit of the TO as a function of the mean value $\mu_p$ for different pricing schemes and values of $\pi_m^{\mathcal{P}}$ (Solid lines: $\pi_m^{\mathcal{P}} = 5 \cdot 10^{-5}$; Dashed lines: $\pi_m^{\mathcal{P}} = 10^{-2}$).

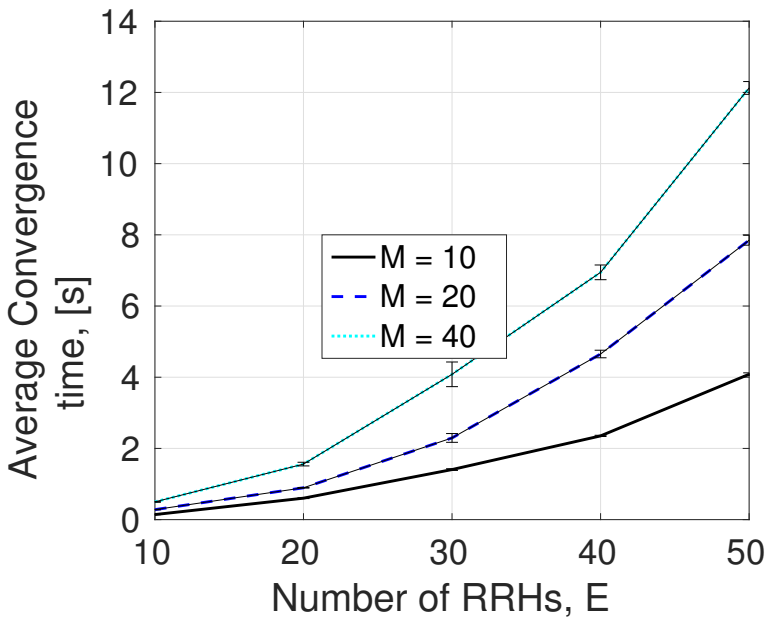

Figure 13. Experimental results of convergence time as a function of the number $R$ or RRHs for different values of the number $M$ of MVNOs.

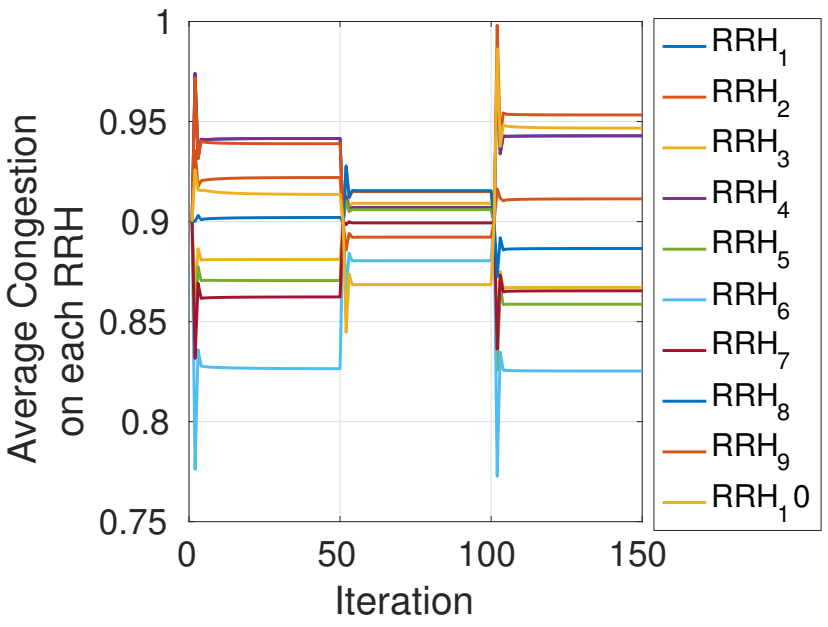

Figure 14. Dynamic behavior of the proposed slicing algorithm.

Fig. 13, the learning algorithm reaches a NE in a limited amount of time (less that 13 seconds for 50 RRHs and 40 MVNOs), despite the overhead due to communication between the TO and the MVNOs.

In Fig. 14, we show how the proposed RAN slicing algorithm adapts to network changes over time when $M = 3$ and $R = 10$. We assume that the number $n_m(t)$ of MUs served by MVNO $m$ and their position changes over time. Specifically, every 50 iterations of Algorithm 2, we randomly generate a new network configuration by updating both the number of MUs and their expected position in the cluster. Accordingly, Fig. 14 shows the average congestion level on each RRH in the network at each iteration of Algorithm 2. It is worth noting that Algorithm 2 quickly adapts to network changes and rapidly converges towards a NE in few iterations even though the network configuration changes over time.

## IX. Related Work

Due to its capability to enable agile and efficient sharing of wireless network resources in 5G systems, the network slicing problem has attracted increasing attention in the literature [10, 13, 35–37]. Network slicing is not trivial, and it can be shown that in general it is NP-hard [10]. Centralized solutions have been proposed for both the slicing of both the backbone and RAN portions of the network [10, 13, 38, 39], but they generally require perfect knowledge of the network configuration and suffers from high complexity issues. As a consequence, a variety of low-complexity heuristic and distributed approaches have been proposed [40–42]. However, the above solutions do not consider the adversarial behavior of virtual operators, which makes it hard to implement them in competitive scenarios where privacy must be preserved among multiple entities.

It is worth noting that many instances of the network slicing problem can actually be reduced to the so-called Virtual Network Embedding (VNE) problem [43], which consists in the problem of generating a set of virtual networks upon a shared physical infrastructure while satisfying a given set of constraints. Unfortunately, the VNE problem is NP-hard. Thus, to design effective algorithms with low computational complexity, a variety of centralized and distributed heuristic approaches have been proposed in the literature [43]. However, all of those solutions are not well-suited to deal with the competitive slicing of the RAN, where information on the expected number of users and their distribution in the network is available. Thus, they cannot be straightforwardly applied to the RAN slicing problem we are tackling in this paper.

Competition among virtual operators has been recently considered in [25], where game-theoretical tools have been effectively exploited to address the RAN slicing problem. In [25], a RAN slicing game is developed, where the available RAN resources are allocated to the MUs through an auction-based mechanism which competitively maximizes the data rate of each slice while guaranteeing a certain level of fairness.

Though the problem tackled in [25] is similar to that addressed in this paper, there are several fundamental differences. First, in [25] a fixed share of RAN resources is assigned to each MVNO a-priori, and the auction game is then played by the MVNOs to decide on how to divide those resources to the mobile users. In our paper, conversely, the share of RAN resources allocated to each MVNO has been obtained as a consequence of a non-cooperative game played among the competing MVNOs. Another important difference is that perfect knowledge about the presence of users is assumed in [25]. In our paper, instead, we have relaxed this assumption by assuming that only statistical information on the presence and position of the users is available.

In the previous sections, we have emphasized the importance of deriving theoretical bounds on the efficiency of the NE, and designing effective distributed and convergent algorithms. Due to their interesting properties with respect to the above issues, congestion games [16], and their application to many networking problems such as the network service chaining [15] and load balancing [44], have been investigated in the literature [14, 26, 28, 45–47].

In this paper, we have leveraged on a particular class of congestion games, i.e., the atomic splittable congestion games. Although the uniqueness of the NE for this class of games is a well-known result in the literature, how to efficiently compute a NE is still a challenging issue. Closed-form solutions have been derived in [45], which however only hold for resource-specific cost functions. When closed-form solutions cannot be derived, then traditional sequential BRD have been considered [46]. However, those approaches generally require to solve QP problems, and are guaranteed to converge to the unique NE only after an infinite number of iterations. Recently, a quantization approach which can compute the NE in a finite amount of time has been proposed in [26]. However, it suffers from highly polynomial complexity, which makes it hard to implement it in dynamic scenarios. Upper-bounds on the PoA of atomic congestion games have been investigated in [28] and [47], but they only apply to the case of resource-specific cost functions.

If compared to the above literature on atomic congestion games, in this work we have proposed a simple low-complexity learning algorithm where each MVNO iteratively updates its strategy to provably converge to the unique NE. We have further derived an upper-bound on the PoA of the class of atomic splittable congestion games where resource and player-specific cost functions are considered, thus extending the results in [28] and [47]. Specifically, we have proved that the PoA for this particular class of cost functions matches the PoA of atomic splittable congestion games where only resource-specific cost functions are considered [28].

## X. Conclusions

In this paper, we have leveraged congestion games and learning mechanisms to design a distributed solution to the wireless network slicing problem. The proposed solution accounts for the limited availability of wireless resources and considers several aspects such as congestion, deployment costs and distance among the RRHs and MUs. We have shown that the proposed solution achieves near-optimal global performance with limited computational complexity while preserving the privacy of MVNOs. Numerical and experimental results have shown the effectiveness of the proposed approach in terms of cost reduction, scalability and convergence time.

## Appendix

### A. Proof of Proposition 3 (Potential Function and NE Uniqueness)

*Proof:* The cost functions in (11) are strictly convex in each player's strategy $\boldsymbol{\xi}_m$. Thus, the general theory of [48] on concave games ensures the existence of at least one NE.

To prove that $\mathcal{G}$ is an exact potential game, we must show that $c_m(\boldsymbol{\xi}_m, \boldsymbol{\xi}_{-m}) - c_m(\boldsymbol{\xi}'_m, \boldsymbol{\xi}_{-m}) = \Phi(\boldsymbol{\xi}_m, \boldsymbol{\xi}_{-m}) - \Phi(\boldsymbol{\xi}'_m, \boldsymbol{\xi}_{-m})$ for all $m \in \mathcal{M}$ and $(\boldsymbol{\xi}_m, \boldsymbol{\xi}_{-m}), (\boldsymbol{\xi}'_m, \boldsymbol{\xi}_{-m}) \in \mathcal{S}$.

Let us consider the $j$-th MVNO in $\mathcal{M}$, and consider two allocation profiles $(\boldsymbol{\xi}_j, \boldsymbol{\xi}_{-j})$ and $(\boldsymbol{\nu}_j, \boldsymbol{\nu}_{-j})$ such that $\boldsymbol{\xi}_{-j} = \boldsymbol{\nu}_{-j}$ and $\boldsymbol{\xi}_j \neq \boldsymbol{\nu}_j$. Since $\boldsymbol{\xi}_{-j} = \boldsymbol{\nu}_{-j}$, we have

$$\Phi(\boldsymbol{\xi}_j, \boldsymbol{\xi}_{-j}) - \Phi(\boldsymbol{\nu}_j, \boldsymbol{\xi}_{-j}) = \sum_{r\in\mathcal{R}} \frac{1}{N_r}(\xi_{j,r}^2 - \nu_{j,r}^2) + \eta_{j,r}(\xi_{j,r} - \nu_{j,r}) + \sum_{r\in\mathcal{R}} \sum_{m\in\mathcal{M}} \frac{1}{N_r}\left(\xi_{m,r}\sum_{k<m}\xi_{k,r} - \nu_{m,r}\sum_{k<m}\nu_{k,r}\right) \quad (29)$$

Note that

$$\sum_{m\in\mathcal{M}} \xi_{m,r}\sum_{k<m}\xi_{k,r} = \sum_{m=1}^{j-1}\xi_{m,r}\sum_{k<m}\xi_{k,r} + \sum_{m=j}^{M}\xi_{m,r}\sum_{k<m}\xi_{k,r} \quad (30)$$

Since $\boldsymbol{\xi}_{-j} = \boldsymbol{\nu}_{-j}$ by assumption, we have that $\sum_{m=1}^{j-1}\xi_{m,r}\sum_{k<m}\xi_{k,r} = \sum_{m=1}^{j-1}\nu_{m,r}\sum_{k<m}\nu_{k,r}$. Thus, the last summation in (29) can be rewritten as follows:

$$\begin{aligned}
&\sum_{r\in\mathcal{R}} \sum_{m\in\mathcal{M}} \frac{1}{N_r}\left(\xi_{m,r}\sum_{k<m}\xi_{k,r} - \nu_{m,r}\sum_{k<m}\nu_{k,r}\right) = \\
&\sum_{r\in\mathcal{R}} \frac{1}{N_r}\left(\xi_{j,r}\sum_{k=1}^{j-1}\xi_{k,r} - \nu_{j,r}\sum_{k=1}^{j-1}\nu_{k,r}\right) \\
&+ \sum_{r\in\mathcal{R}} \frac{1}{N_r}\sum_{m=j+1}^{M}\left(\xi_{m,r}\sum_{k<m}\xi_{k,r} - \nu_{m,r}\sum_{k<m}\nu_{k,r}\right) = \\
&\sum_{r\in\mathcal{R}} \frac{1}{N_r}\left(\xi_{j,r}\sum_{k=1}^{j-1}\xi_{k,r} - \nu_{j,r}\sum_{k=1}^{j-1}\nu_{k,r}\right) \\
&+ \sum_{r\in\mathcal{R}} \frac{1}{N_r}\sum_{m=j+1}^{M}\xi_{m,r}\left(\xi_{j,r} - \nu_{j,r}\right) = \\
&\sum_{r\in\mathcal{R}} \frac{1}{N_r}\left[\xi_{j,r}\sum_{m\neq j}\xi_{m,r} - \nu_{j,r}\sum_{m\neq j}\xi_{m,r}\right] \quad (31)
\end{aligned}$$

where the last step is obtained by exploiting $\boldsymbol{\xi}_{-j} = \boldsymbol{\nu}_{-j}$. By substituting (31) in (29), we obtain:

$$\begin{aligned}
&\Phi(\boldsymbol{\xi}_j, \boldsymbol{\xi}_{-j}) - \Phi(\boldsymbol{\xi}_j, \boldsymbol{\xi}_{-j}) = \sum_{r\in\mathcal{R}} \frac{1}{N_r}\left(\xi_{j,r}^2 + \xi_{j,r}\sum_{m\neq j}\xi_{m,r}\right) + \\
&\eta_{j,r}\xi_{j,r} - \sum_{r\in\mathcal{R}} \frac{1}{N_r}\left(\nu_{j,r}^2 + \nu_{j,r}\sum_{m\neq j}\nu_{m,r}\right) + \eta_{j,r}\nu_{j,r} = \\
&c_m(\boldsymbol{\xi}_j, \boldsymbol{\xi}_{-j}) - c_m(\boldsymbol{\nu}_j, \boldsymbol{\xi}_{-j}) \quad (32)
\end{aligned}$$

which proves that the function $\Phi(\cdot)$ in (16) is an exact potential function for the game $\mathcal{G}$. Note that the Hessian matrix of $\Phi(\boldsymbol{\xi})$ with respect to $\boldsymbol{\xi}$ has strictly positive eigenvalues, which guarantees that the potential function is strictly convex and admits a unique global minimizer. It is worth noting that $\Phi(\boldsymbol{\xi})$ is a potential function for $\mathcal{G}$, and is convex on the convex set $\mathcal{S}$. Hence, the set of minimizers of $\Phi(\boldsymbol{\xi})$ coincides with the set of NEs of $\mathcal{G}$ [49]. Since $\Phi(\boldsymbol{\xi})$ admits a unique global minimizer, it must follows that the NE of $\mathcal{G}$ is unique. ■

### B. Proof of Proposition 4 (Bound on the PoA)

*Proof:* Let us consider two allocation profiles $\boldsymbol{\xi}, \boldsymbol{\nu} \in \mathcal{S}$ such that $\boldsymbol{\xi}$ is the unique NE of game $\mathcal{G}$, and $\boldsymbol{\nu}$ is the optimal solution of (17). From (18) and (11), we have

$$\begin{aligned}
C^{\text{NE}} &= \sum_{m\in\mathcal{M}}\sum_{r\in\mathcal{R}}\xi_{m,r}c_{m,r}(\boldsymbol{\xi}) = \\
&\sum_{m\in\mathcal{M}}\sum_{r\in\mathcal{R}}\xi_{m,r}\left(c_{m,r}(\boldsymbol{\xi}) - v_{m,r}(\boldsymbol{\xi}) + v_{m,r}(\boldsymbol{\xi})\right) \quad (33)
\end{aligned}$$

Since $\xi_{m,r}c_{m,r}(\boldsymbol{\xi})$ is convex in $\boldsymbol{\xi}$, we have that $v_{m,r}(\boldsymbol{\xi})(\nu_{m,r} - \xi_{m,r}) \geq 0$. Hence, by exploiting (19), (33) can be maximized as

follows:

$$C^{\mathrm{NE}} \leq \sum_{m\in\mathcal{M}}\sum_{r\in\mathcal{R}} \xi_{m,r}(c_{m,r}(\boldsymbol{\xi}) - \nu_{m,r}(\boldsymbol{\xi})) + \nu_{m,r}\nu_{m,r}(\boldsymbol{\xi}) = \sum_{m\in\mathcal{M}}\sum_{r\in\mathcal{R}} \xi_{m,r}(c_{m,r}(\boldsymbol{\xi}) - \nu_{m,r}(\boldsymbol{\xi})) + \nu_{m,r}(\nu_{m,r}(\boldsymbol{\xi}) - c_{m,r}(\boldsymbol{\nu}) + c_{m,r}(\boldsymbol{\nu})) = \sum_{m\in\mathcal{M}}\sum_{r\in\mathcal{R}} \left[\xi_{m,r}(c_{m,r}(\boldsymbol{\xi}) - \nu_{m,r}(\boldsymbol{\xi})) + \nu_{m,r}(\nu_{m,r}(\boldsymbol{\xi}) - c_{m,r}(\boldsymbol{\nu}))\right] + C^{\mathrm{OPT}} \leq \beta \sum_{r\in\mathcal{R}}\sum_{m\in\mathcal{M}} \xi_{m,r}c_{m,r}(\boldsymbol{\xi}) + C^{\mathrm{OPT}} = \beta C^{\mathrm{NE}} + C^{\mathrm{OPT}} \quad (34)$$

which implies that $C^{\mathrm{NE}} \leq \frac{1}{1-\beta} \cdot C^{\mathrm{OPT}}$. ■

### C. Proof of Theorem 5 (The PoA is upper bounded)

*Proof:* Let us consider two allocation profiles $\boldsymbol{\xi}, \boldsymbol{\nu} \in \mathcal{S}$. From (20), the numerator of (19) can be rewritten as

$$\sum_{m\in\mathcal{M}} \xi_{m,r}\left(c_{m,r}(\boldsymbol{\xi}) - \nu_{m,r}(\boldsymbol{\xi})\right) + \sum_{m\in\mathcal{M}} \nu_{m,r}\left(\nu_{m,r}(\boldsymbol{\xi}) - c_{m,r}(\boldsymbol{\nu})\right) = \sum_{m\in\mathcal{M}} \xi_{m,r}c_{m,r}(\boldsymbol{\xi}) - \sum_{m\in\mathcal{M}} \nu_{m,r}c_{m,r}(\boldsymbol{\nu}) + \sum_{m\in\mathcal{M}} \nu_{m,r}(\boldsymbol{\xi})\left(\nu_{m,r} - \xi_{m,r}\right) = \sum_{m\in\mathcal{M}} \nu_{m,r}\left(c_{m,r}(\boldsymbol{\xi}) - c_{m,r}(\boldsymbol{\nu})\right) + \frac{1}{N_r}\sum_{m\in\mathcal{M}}\left(\nu_{m,r}\xi_{m,r} - \xi^2_{m,r}\right) = \sum_{m\in\mathcal{M}} \nu_{m,r}\left(c_{m,r}(\boldsymbol{\xi}) - c_{m,r}(\boldsymbol{\nu}) + \frac{1}{N_r}\xi_{m,r}\right) - \frac{1}{N_r}\xi^2_{m,r} \quad (35)$$

For each $r \in \mathcal{R}$, let $Y_r = \sum \nu_{m,r}$ and $X_r = \sum_{m\in\mathcal{M}} \xi_{m,r}$, where $\xi_{m,r} \in \boldsymbol{\xi}$ and $\nu_{m,r} \in \boldsymbol{\xi}$. From (9) and (10), we have that

$$c_{m,r}(\boldsymbol{\xi}) = \phi_r(\boldsymbol{\xi}) + \eta_{m,r} = \frac{X_r}{N_r} + \eta_{m,r} \quad (36)$$

and

$$c_{m,r}(\boldsymbol{\xi}) - c_{m,r}(\boldsymbol{\nu}) = \frac{1}{N_r}(X_r - Y_r) \quad (37)$$

From (36) and (37), (35) can be reformulated as follows:

$$\sum_{m\in\mathcal{M}} \nu_{m,r}\left(c_{m,r}(\boldsymbol{\xi}) - c_{m,r}(\boldsymbol{\nu}) + \frac{1}{N_r}\xi_{m,r}\right) - \frac{1}{N_r}\xi^2_{m,r} = \frac{1}{N_r}\sum_{m\in\mathcal{M}} \nu_{m,r}(X_r - Y_r + \xi_{m,r}) - \frac{1}{N_r}\sum_{m\in\mathcal{M}}\xi^2_{m,r} = \frac{1}{N_r}(X_r - Y_r)Y_r + \frac{1}{N_r}\sum_{m\in\mathcal{M}}\nu_{m,r}\xi_{m,r} - \frac{1}{N_r}\sum_{m\in\mathcal{M}}\xi^2_{m,r} \leq \frac{1}{N_r}\sup_{k\in\mathcal{M}}\left\{(X_r - Y_r + \xi_{k,r})Y_r - \sum_{m\in\mathcal{M}}\xi^2_{m,r}\right\} \quad (38)$$

Note that

$$Y_r(X_r - Y_r + \xi_{k,r}) = \frac{(X_r + \xi_{k,r})^2}{4} - \left(\frac{X_r + \xi_{k,r}}{2} - Y_r\right)^2 \leq \frac{(X_r + \xi_{k,r})^2}{4} \quad (39)$$

Thus, the last term in (38) can be further maximized as follows:

$$\sup_{k\in\mathcal{M}}\left\{(X_r - Y_r + \xi_{k,r})Y_r - \sum_{m\in\mathcal{M}}\xi^2_{m,r}\right\} \leq \sup_{k\in\mathcal{M}}\left\{\frac{(X_r + \xi_{k,r})^2}{4} - \sum_{m\in\mathcal{M}}\xi^2_{m,r}\right\} \quad (40)$$

It is worth noting that (40), which is a maximization of the numerator of $\beta$ in (19), does not depend on the value of $\boldsymbol{\nu}$ anymore. Accordingly, we have that

$$\beta \leq \frac{1}{4}\sup_{r\in\mathcal{R}}\frac{1}{N_r}\sup_{\boldsymbol{\xi}\in\mathcal{S},k\in\mathcal{M}}\left\{\frac{(X_r + \xi_{k,r})^2 - 4\sum_{m\in\mathcal{M}}\xi^2_{m,r}}{\sum_{m\in\mathcal{M}}\xi_{m,r}c_{m,r}(\boldsymbol{\xi})}\right\} \leq \frac{1}{4}\sup_{r\in\mathcal{R}}\frac{1}{N_r}\sup_{\boldsymbol{\xi}\in\mathcal{S},k\in\mathcal{M}}\left\{\frac{X_r^2 + 2\xi_{k,r}X_r - 3\xi^2_{k,r} - 4\sum_{m\neq k}\xi^2_{m,r}}{\sum_{m\in\mathcal{M}}\xi_{m,r}c_{m,r}(\boldsymbol{\xi})}\right\} \quad (41)$$

where we recall that $\xi_{k,r} \in \boldsymbol{\xi}$, and $X_r = \sum_{m\in\mathcal{M}}\xi_{m,r}$ with $\xi_{m,r} \in \boldsymbol{\xi}$.

By substituting (36) in the denominator of (41), and by maximizing it by removing the term $\eta_{m,r}$, we obtain

$$\beta \leq \frac{1}{4}\sup_{r\in\mathcal{R}}\frac{1}{N_r}\sup_{\boldsymbol{\xi}\in\mathcal{S},k\in\mathcal{M}}\left\{\frac{X_r^2 + 2\xi_{k,r}X_r - 3\xi^2_{k,r} - 4\sum_{m\neq k}\xi^2_{m,r}}{\frac{1}{N_r}X_r\sum_{m\in\mathcal{M}}\xi_{m,r}}\right\} \leq \frac{1}{4}\sup_{r\in\mathcal{R}}\sup_{\boldsymbol{\xi}\in\mathcal{S},k\in\mathcal{M}}\left\{\frac{X_r^2 + 2\xi_{k,r}X_r - 3\xi^2_{k,r} - 4\sum_{m\neq k}\xi^2_{m,r}}{X_r^2}\right\} \leq \frac{1}{4}\sup_{\mathbf{t}\in\mathcal{T}}\left\{1 + 2t_k - 3t_k^2 - 4\sum_{m\neq k}t_m^2\right\} \quad (42)$$

where $t_m = \frac{\xi_{m,r}}{X_r}$, $\mathbf{t} = (t_m)_{m\in\mathcal{M}}$, and $\mathcal{T} = \{t_m \in [0,1] : \sum_{m\in\mathcal{M}} t_m = 1, t_k \geq t_j \ \forall j \in \mathcal{M}\}$. Since $X_r = \sum_{m\in\mathcal{M}}\xi_{m,r}$, we have that (42) does not depend on $r \in \mathcal{R}$ anymore. Also, let us note that the supremum of the function in the r.h.s. of (42) is achieved when

$$t_k = \frac{3+M}{1+3M} \text{ and } t_j = \frac{2}{1+3M} \ \forall j \in \mathcal{M} \setminus k \quad (43)$$

By plugging (43) in (42), we obtain

$$\beta \leq \frac{1}{4}\left[\frac{12M^2 - 8M - 4}{(1+3M)^2}\right] = \frac{M-1}{1+3M} \quad (44)$$

From (44) and Proposition 4, we obtain that $\frac{C^{\mathrm{NE}}}{C^{\mathrm{OPT}}} \leq \frac{1}{1-\beta} \leq \frac{3M+1}{2M+2}$, which concludes the proof. ■